%% file: ms.tex
\documentclass[letterpaper,twocolumn,10pt]{article}
\usepackage{usenix,epsfig,endnotes}
\usepackage[compact]{titlesec}
\usepackage{titlesec}

\usepackage{colortbl}
\usepackage{verbatim}
\usepackage{color}
\usepackage{balance}
\usepackage{tikz}
\usepackage{cite}
\usepackage{enumitem}
\usepackage{subfig}
\usepackage{multirow}
\usepackage{array}
\usepackage{tabularx}
\usepackage{pifont}
\usepackage[hyphens]{url}
\definecolor{darkgreen}{rgb}{0.00,0.39,0.00}
\usepackage[
    colorlinks,
    linkcolor=blue,
    citecolor=darkgreen,
    urlcolor=blue,
]{hyperref}
\usepackage{abbrevs}
\usepackage{xspace} 
\definecolor{red3}{rgb}{0.80,0.00,0.00}
\newcommand{\note}[1]{[\textcolor{red}{\textit{#1}}]}

\newcommand{\IKO}[1]{\note{IKO: #1}}

\newcommand{\Sec}[1]{\S{\ref{#1}}}
\newcommand{\Fig}[1]{Fig.~\ref{#1}}

\newcommand{\Tab}[1]{Table~\ref{#1}}

\newcommand{\pr}[1]{{\small\texttt{#1}}}

\makeatletter
\let\maybe@space@\xspace
\makeatother

\newabbrev\SCM{storage capacity manager (SCM)}[SCM]
\newabbrev\GC{garbage collection (GC)}[GC]
\newabbrev\SMR{Shingled Magnetic Recording (SMR)}[SMR]
\newabbrev\SSD{Solid State Drives (SSDs)}[SSD]
\newcommand{\Sys}{SALSA\xspace}

\providecommand{\tightlist}{%
  \setlength{\itemsep}{0pt}\setlength{\parskip}{0pt}}

\newcommand{\CM}[1]{}

\begin{document}

\title{Elevating commodity storage with the \Sys host translation layer}
\author{
Nikolas Ioannou,
Kornilios Kourtis, and
Ioannis Koltsidas\\
IBM Research, Zurich\\
\textit{\{nio, kou, iko\}@zurich.ibm.com}
} 

\maketitle

\newcommand\copyrighttext{%
  \footnotesize Published in 2018 IEEE 26th International Symposium on Modeling, Analysis, and Simulation of Computer and Telecommunication Systems (MASCOTS)}
\newcommand\copyrightnotice{%
\begin{tikzpicture}[remember picture,overlay]
\node[anchor=south,yshift=10pt] at (current page.south) {\copyrighttext};
\end{tikzpicture}%
}
\copyrightnotice

\sloppy
\input{abstract}
\input{introduction}

\input{motivation}
\input{system}
\input{evaluation}

\input{related}
\input{conclusion}

\bigskip
\noindent\textbf{Notes:} IBM is a trademark of International
Business Machines Corporation, registered in many jurisdictions worldwide. 
Linux is a registered trademark of Linus Torvalds in the United States, 
other countries, or both.  Other products and service names might be 
trademarks of IBM or other companies.
\bibliographystyle{acm}
\balance
\bibliography{ms}

\end{document}

%% file: abstract.tex
\begin{abstract}

To satisfy increasing storage demands in both capacity and performance, industry has turned to multiple storage technologies, including Flash SSDs and SMR disks. These devices employ a translation layer that conceals the idiosyncrasies of their mediums and enables random access. Device translation layers are, however, inherently constrained: resources on the drive are scarce, they cannot be adapted to application requirements, and lack visibility across multiple devices. As a result, performance and durability of many storage devices is severely degraded.

In this paper, we present SALSA\@: a translation layer that executes on the
host and allows unmodified applications to better utilize commodity storage.
SALSA  supports a wide range of single- and multi-device optimizations
and, because is implemented in software, can adapt to specific
workloads. We describe SALSA's design, and demonstrate its significant benefits
using microbenchmarks and case studies based on three applications:
MySQL, the Swift object store, and a video server.

\end{abstract}

%% file: introduction.tex
\section{Introduction}\label{introduction}

The storage landscape is increasingly diverse. The market is
dominated by spinning magnetic disks (HDDs) and Solid State Drives (SSDs) based
on NAND Flash.
Broadly speaking, HDDs offer lower cost per GB, while
SSDs offer better performance, especially for read-dominated workloads.
Furthermore, emerging technologies provide new tradeoffs: Shingled Magnetic Recording (SMR)
disks\cite{wood:tm09} offer increased capacity compared to HDDs,
while Non-Volatile Memories (NVM)\cite{nanavati15:nvm} offer persistence with
performance characteristics close to that of DRAM.
At the same time, applications have different requirements and
access patterns, and no one-size-fits-all storage solution exists.
Choosing between SSDs, HDDs, and SMRs, for example,
depends on the capacity, performance and cost requirements, as well as on the
workload. To complicate things further, many applications
(e.g., media services\cite{ripq:fast15, google-disks-dc:2016})
require multiple storage media to meet their requirements.

Storage devices are also frequently idiosyncratic. NAND Flash, for example, has a
different access and erase granularity, while SMR disks preclude in-place
updates, allowing only appends. Because upper layers (e.g., databases and
filesystems) are often not equipped to deal with these
idiosyncrasies, translation layers\cite{chung09:ftl} are introduced to enable
applications to access idiosyncratic storage transparently.
Translation layers (TLs) implement an indirection between the logical
space as seen by the application, and the physical storage as exposed by the
device. 
A TL can either be
implemented on the host (host TL) or on the device controller (drive TL).
%
It is well established that, for many workloads, drive TLs lead to sub-optimal use
of the storage medium.
Many works identify these performance problems, and try to address them by
improving the controller translation layer\cite{ma14:ftl-survey,gupta09:dftl},
or adapting various layers of the I/O software stack:
filesystems\cite{josephson:2010:dfs,min:2012:sfs,f2fs:2015},
caches\cite{park06:cflru,kim08:bplru}, paging\cite{saxena10:fvm}, and key-value
stores\cite{debnath10:flashstore,debnath11:skimpystash,marmol14:nvmkv,wang14:lsm-ssd,pitchumani15:smrdb}.



In agreement with a number of recent
works\cite{ouyang14:sdf,lightnvme:fast17,amf:fast16}, we argue that
these shortcomings are inherent to drive TLs,
and advocate placing
the TL on the host. While a host TL is not a new
idea\cite{fusionio:ftl:2011,object-ftl:fast13}, our approach is different from
previous works in a number of ways.
%
First, we focus on commodity drives, without dependencies on specific vendors.
Our goal is to enable datacenter applications to use cost-effective storage,
while maintaining acceptable performance.
Second, we propose a unified TL framework
that supports different storage technologies (e.g., SSDs, SMRs).
Third, we argue for a host TL that can be adapted to different application
requirements and realize different tradeoffs.
Finally, we propose a TL that can virtualize multiple devices, potentially of
different types.
The latter allows optimizing TL functions such as load
balancing and wear leveling across devices, while also addressing storage
diversity by enabling hybrid
systems that utilize different media.

\Sys (SoftwAre Log Structured Array) implements the above ideas,
following a log-structured architecture\cite{rosenblum92:logfs,menon95:lsa}.
We envision \Sys{} as the backend of a software-defined storage system, where it
manages a shared storage pool, and can be configured to use workload-specific
policies for each application using the storage.
In this paper, we focus on the case where \Sys is used to run unmodified
applications by exposing a Linux block device that can be either used directly,
or mounted by a traditional Linux filesystem.
The contribution of our work is a novel host TL architecture and
implementation that supports different media and allows optimizing for
different objectives.
Specifically:
\begin{itemize}
	\tightlist{}
     \item \Sys{} achieves substantial performance and durability benefits by implementing the
     TL on the host for single- and multi-device setups. When deploying MySQL
     database containers on commodity SSDs, \Sys{} outperforms the raw device by
     $1.7 \times$ on one SSD and by $35.4 \times$ on a software RAID-5
     array.

    \item \Sys{} makes efficient use of storage by allowing application-specific
    policies. We present a \Sys policy tailored to the Swift object
    store\cite{openstack-swift} on SMRs that outperforms the raw device by up to
    a factor of $6.3 \times$.

    \item \Sys{} decouples space management from storage policy. This enables \Sys{} to
    accommodate different applications, each with its own policy, using the same
    storage pool. This allows running MySQL and Swift on the same storage with
    high performance and a low overhead.

    \item \Sys{} embraces storage diversity by supporting multiple types of devices. We
    present how we combine SMRs and SSDs to speedup file retrieval for a video
    server workload (where adding an SSD improves read performance by $19.1
    \times$), without modifying the application.

\end{itemize}

The remaining of the paper is organized as follows. We start with our a brief
overview of idiosyncratic storage and our motivation behind \Sys
(\Sec{sec:motivation}). We continue with a description of the design of \Sys
(\Sec{sec:system}), discuss how we satisfy specific application
workload requirements (\Sec{sec:adaptation}), and evaluate our approach
(\Sec{sec:evaluation}). Finally, we discuss related work (\Sec{sec:related}) and
conclude (\Sec{sec:conclusions}).

%% file: motivation.tex
\section{Background and Motivation}
\label{sec:motivation}

In this section, we provide a brief background on Flash-based SSDs
(\Sec{ssec:bg:flash}) and \SMR disks (\Sec{ssec:bg:smr-disks}), analyze the
limitations of commodity drive TLs (\Sec{ssec:bg:ftl-limitations}), and argue
for a unified host TL architecture (\Sec{ssec:bg:host-tl}).

\subsection{Flash-based SSDs}
\label{ssec:bg:flash}


Flash memory fits well in the gap between DRAM and spinning disks: it offers
low-latency random accesses compared to disks at a significantly lower
cost than DRAM.  As a result, its adoption is constantly increasing in the data
center\cite{google:flash:fast16, idc-flash, knipple:flash17}, where it is
primarily deployed in the form of SSDs.
%
%
Nevertheless, Flash has unique characteristics that complicate its
use\cite{andersen10:flash-dc}.
First, writes are significantly more involved than reads. NAND Flash memory is
organized in pages, and a page needs to be \emph{erased} before it can be
\emph{programmed} (i.e., set to a new value). Not only programming a page is
much slower than reading it, but the erase operation needs to be performed in
blocks of (hundreds or even thousands of) pages.  Therefore, writes
cannot be done in-place, and also involve a high cost as block erasures are two
orders of magnitude slower than reading or programming a page.
Second, each Flash cell can only sustain a finite number of erase cycles before
it wears out and becomes unusable.
%


\label{ssec:bg:ftl}

Flash translation layers (FTLs)\cite{ma14:ftl-survey} are introduced to address
the above issues.  In general, an FTL performs writes out-of-place, and maintains
a mapping between logical and physical addresses. When space runs out, invalid
data are garbage collected, and valid data are relocated to free blocks. To aid
the \GC process, controllers keep a part of the drive's capacity hidden from the
user (\emph{overprovisioning}).  The more space that is overprovisioned, the better
the \GC performs.

\subsection{SMR disks}
\label{ssec:bg:smr-disks}


Magnetic disks remain the medium of choice for many
applications\cite{google-disks-dc:2016}, mainly due to low cost.
However, magnetic recording is reaching its density scaling limits. To increase
disk capacity, a number of techniques have been proposed, one of which, Shingled
Magnetic Recording (SMR) \cite{feldman2013shingled,wood:tm09}, has recently become
widely available\cite{st8:manual,ha10:manual}.
SMRs gain density by precluding random updates. The density improvement is
achieved by reducing the track width, thereby fitting more tracks on the surface
of a platter, without reducing the write head size. As a result, while a track
can be read as in conventional disks, it cannot be re-written without damaging
adjacent tracks.

SMR disks are typically organized into zones. Zones are isolated from each other
by guards, so that writes to a zone do not interfere with tracks on
other zones. All the writes within a zone must be done in strict sequential
order; however, multiple zones can be written to independently. These zones are
called \emph{sequential zones}. Drives, also, typically include a small number of
\emph{conventional zones}, where random updates are allowed.

Three categories of SMR drives exist based on where the TL
is placed: drive-managed (DM), host-managed (HM), and host-aware
(HA) SMRs\cite{t10-zbc}. In drive-managed disks, SMR complexities are fully
hidden by a drive TL\@.
On the other extreme, HM drives
\emph{require} a host TL to guarantee that writes within a zone will be
sequential. HM drives provide a number of commands to the host, e.g., to reset a
zone so that it can be re-written and to retrieve the drive's zone information.
HA SMRs offer a compromise between DM and HM: they
expose control commands, but can operate without a host TL.

\subsection{Limitations of drive TLs}
\label{ssec:bg:ftl-limitations}


Demand to reduce costs per GB raises barriers to drive TL performance.
SSD vendors increasingly offer commodity drives of higher
densities at lower prices, without adding hardware resources (e.g., memory and
compute) on the controller to deal with the extra capacity. Performance degrades further due 
to the use of consumer-grade Flash and low
overprovisioning, as is typical in commodity SSDs. Furthermore, drive TLs are
required to support a number of different workloads, and end
up making compromises. Communicating application-specific hints to the drive is hard, if not
impossible.

\begin{figure}
    \subfloat[Write Bandwidth]{%
        \includegraphics[width=.48\linewidth]{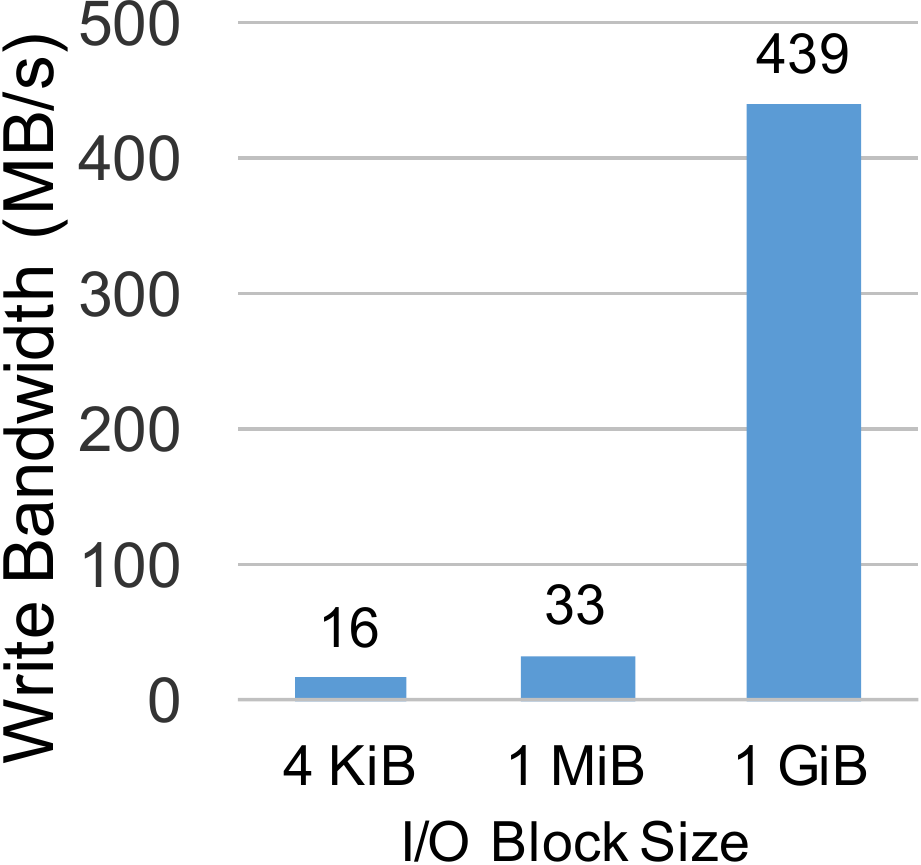}
        \label{fig:bg:ubench:ssd-perf}
    }
    \subfloat[Wear-out after 10 device writes] {%
        \includegraphics[width=.48\linewidth]{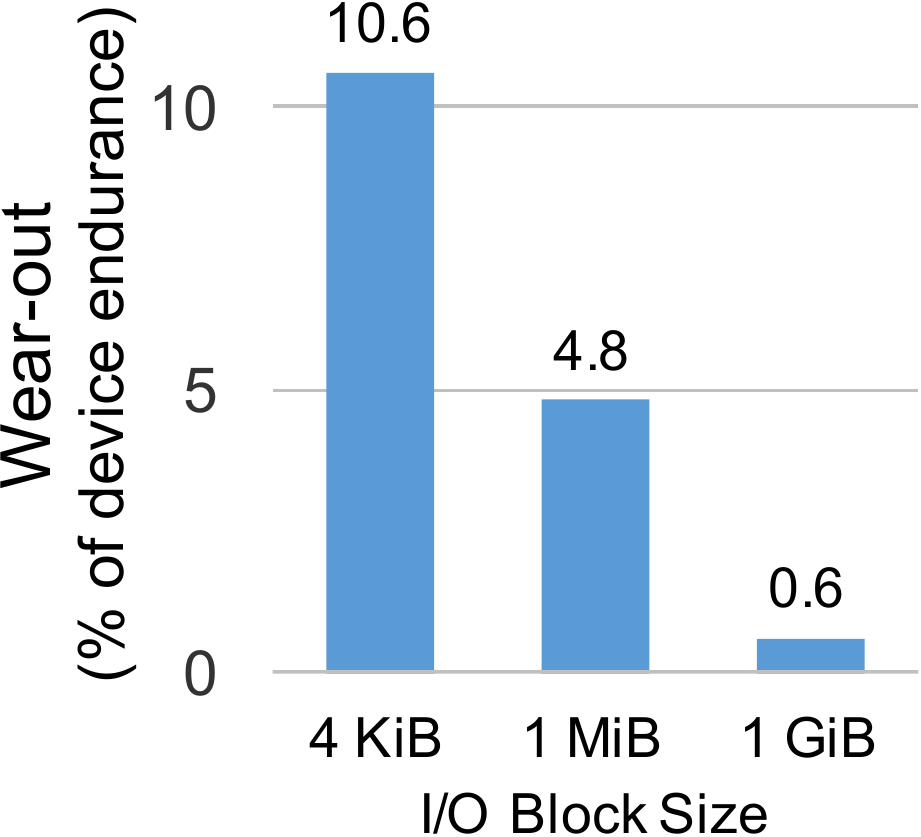}
        \label{fig:bg:ubench:ssd-endurance}
    }
    \caption{Random writes with a block size of 4KiB, 1MiB, and 1GiB}
    \label{fig::bg:ubench}
\end{figure}

We illustrate the limitations of drive TLs on commodity SSDs by applying
a random workload on a widely-used drive (after low-level formatting it) for tens of
hours until performance stabilizes (see \Sec{ssec:ubench:ssd} for
details). We perform the experiment for three block sizes: 4KiB, 1MiB, and 1GiB.
\Fig{fig:bg:ubench:ssd-perf} shows the resulting (stable) write throughput, and
\Fig{fig:bg:ubench:ssd-endurance} shows the wear induced to the device (as
reported by SMART) after 10 full device writes.  Our experiment, effectively,
compares the drive TL performance under a random workload (4KiB) versus the
ideal performance (1GiB), as well as an intermediate performance point
(1MiB). A larger block size minimizes the need for the drive TL to perform \GC:
as the I/O size increases so does the probability that a write will entirely
invalidate the Flash blocks it touches, eliminating the need for
relocations. The drive TL fails to achieve high write bandwidth under
unfavourable access patterns, only sustaining about 16MiB/s for 4KiB blocks, and
33MiB/s for 1MiB blocks. Interestingly, a block size of 1MiB is not large
enough to bring the write performance of the drive to its ideal level; block
sizes closer to the GiB level are required instead, which better reflects the
native block size of modern dense Flash
SSDs~\cite{andromeda17nvmw}. Furthermore, according to the SSD's SMART
attributes, the write amplification for the 1GiB writes was $1.03 \times$,
whereas for the 4KiB writes it was $18.24 \times$, and $8.26 \times$ for 1MiB
writes.  We found that other commodity SSDs exhibit similar behavior, with
write amplification factors as high as $50 \times$.  SMR drive TLs suffer from
the same limitations as FTLs.  As an example, we measured less than 200 KiB/s
of random write bandwidth for 64KiB random writes to a drive-managed SMR disk
(\Sec{ssec:ubench:smr}). Overall, there seems to be significant room for
improvement even for a single drive by employing a host TL that does its own
relocations (additional reads and writes), but always writes sequentially to
the device.

\subsection{Why a host TL?}
\label{ssec:bg:host-tl}

Vendors prioritize cost over performance for commodity drives, resulting
in drives that are unfit for many applications that require high
performance in terms of throughput and latency. Even simple
techniques to alleviate this problem (e.g., configurable
overprovisioning) are non-trivial to implement and rarely
applied in practice.

We argue that a host TL can address these issues and
improve efficiency. By transforming the user access pattern to be sequential,
a host TL can realize significant performance and endurance benefits, enabling
commodity drives to be used for datacenter applications even under demanding
performance requirements. Furthermore, having visibility across multiple devices
enables optimizations that are not possible from within a single drive.  An
evaluation of the Aerospike NoSQL store\cite{ocz-aerospike}, for example, has
shown the advantages of managing arrays of Flash SSDs as opposed to individual
drives (e.g., by coordinating \GC cycles across multiple devices).

Moreover, maximizing I/O performance for many application depends on exploiting
workload properties. While this is difficult to do in a device TL, a host TL
offers many such opportunities (e.g., improving performance by
reducing persistence guarantees or sacrificing space). A host TL should be
built as a framework that supports multiple types of storage, different
policies and algorithms, and a wide range of configuration options. A host TL
can, also, be extended and improved over time, allowing incremental adoption of
advanced techniques, new storage technologies, and different tradeoffs.

Finally, and perhaps more importantly, a host TL allows combining multiple
devices to build hybrid storage systems. By managing arrays of devices at the
host, as opposed to a single device in the case of a drive TL, the TL can offer
additional improvements by making global decisions, e.g., about balancing load
and wear across devices. As the storage landscape increasingly diversifies, and
applications require the use of different storage technologies, the existing TL
indirection can be used for implementing hybrid storage policies. Under
this scenario, a host TL is also applicable to technologies that are not, necessarily,
idiosyncratic.

%% file: system.tex
\section{\Sys design}
\label{sec:system}

\Sys makes three principal design choices.
First, it is \emph{log-structured}\cite{rosenblum92:logfs,menon95:lsa}. Among
other benefits, this
allows it to deal with storage idiosyncrasies.  By only writing
big sequential segments, \Sys renders the drive's \GC
irrelevant, as its task becomes trivial.  When space runs out, \Sys
does its own \GC.
Second, \Sys supports multiple storage types,
and can combine them to build hybrid systems.
Finally, \Sys follows a modular design so that it can be used as a framework for
implementing a variety of policies and algorithms, enabling adaptability to
different requirements.

From the perspective of the user, \Sys exposes block devices that can be used
transparently by applications, e.g., by running a database directly on the
device, or by creating a filesystem and running an application on top.
An important benefit of \Sys is that it does
not require any application modifications.

There are two main components in \Sys: the \SCM, which is responsible for
managing the underlying storage, and one or more controllers that operate on the
\SCM (\Fig{fig:salsa-arch}). \SCM is a common substrate that implements storage
provisioning to controllers, \GC, and other common functions.  Controllers are
responsible for implementing the storage policy, performing I/O, and mapping the
logical (application) space to the physical space (\emph{L2P}).



\subsection{Storage Capacity Manager}
\label{ssec:scm}

\begin{figure}[tb]
    \centering
    \includegraphics[width=\columnwidth]{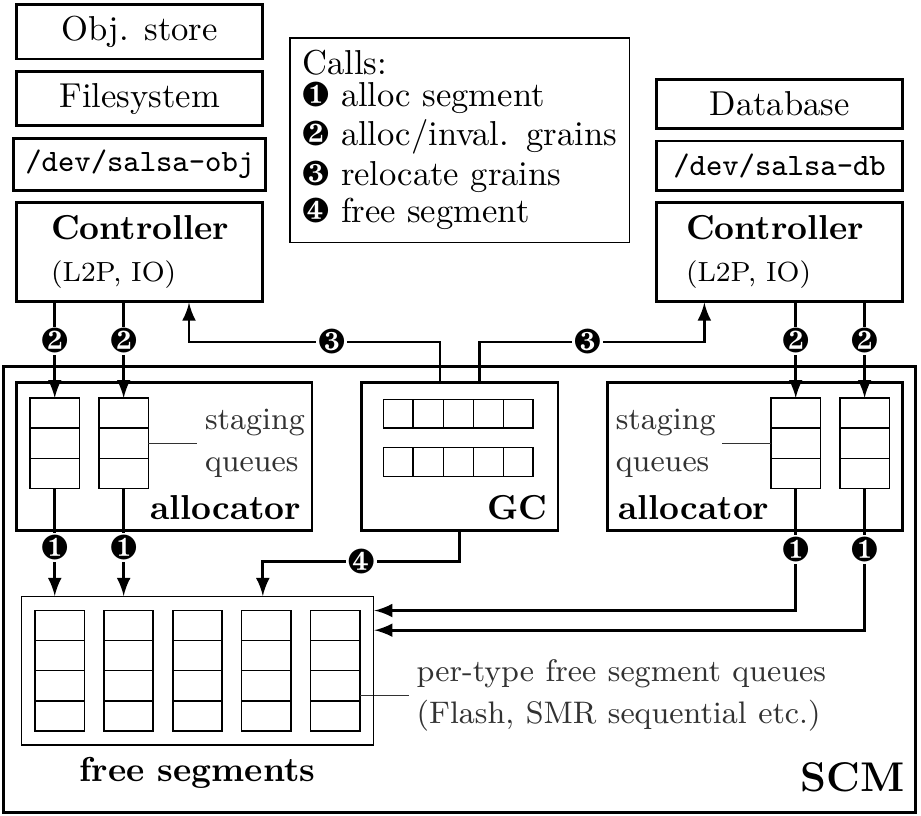}
    \caption{\Sys Architecture and allocation calls.}
    \label{fig:salsa-arch}
\end{figure}

Generally, the \SCM manages multiple storage devices. To capture different device
properties, \Sys defines appropriate storage types: NVM, Flash, HDD, SMR
conventional, SMR sequential. At initialization, \SCM identifies the type
of its devices, and creates an address space that combines them.
This address space is split into different areas, each characterized by a
storage type.
The boundaries of these areas are not necessary device boundaries: an SMR drive
with both conventional and sequential zones, for example, is split into two
different areas, one for each type of zone. The \SCM address space is not
exposed to the controllers.

\Sys can combine multiple devices linearly (i.e., appending one after another)
or in a RAID-0, -1, or \mbox{-5} configuration. Based on the configuration, a set of
appropriate functions for performing I/O is provided to the storage controllers.
For RAID-5, \Sys requires a small non-volatile memory buffer which is used to
store the parity for the currently open stripe. For instance, 
the Persistent Memory Region of an NVMe SSD can be used for that purpose~\cite{nvme1.3}.
Parity is accumulated into that buffer as the stripe is being filled, and is committed 
to storage when it gets full. Thereby, SALSA avoids expensive read-modify-write operations, 
which are required with traditional (i.e., non log-structured) RAID-5.

The \SCM physical storage space is divided into \emph{segments}, large (e.g.,
1GiB) contiguous regions of a single storage type.  A segment can be
in one of the following states: \emph{free}: owned by the \SCM, \emph{staged}:
used for storage allocation for a controller, or \emph{allocated}: fully used,
owned by a controller, and available for \GC. Once allocated, a segment can only
be used by a single controller.

Allocators allocate segments on behalf of the controllers
via an interface (\Fig{fig:salsa-arch}, \ding{182}) that allows for specification of (hard and soft)
constraints on the segment's storage type.  To support this interface, \SCM
maintains multiple allocation queues that segregate segments based on the
backing storage type.
Each segment is divided into \emph{grains}, the configurable smallest unit of
allocation (e.g., 4KiB, 8KiB, etc.). Controllers allocate and free storage in
grains (\ding{183}). Allocators maintain a number of \emph{staged} segments
and allocate space \emph{sequentially} within them. We call this mechanism an
\emph{allocation stream}. Controllers can have multiple allocation streams,
allowing for data segregation. \Sys can, for example, segregate writes by their
update-frequency (``heat''), as well as segregate user and \GC (i.e.,
relocation) writes.  Each stream has its own constraints for segment
allocation.

When a segment becomes full, its state transits to \emph{allocated} and becomes
a relocation candidate for the \GC.
For each segment, \Sys tracks the number
of valid grains. Initially, all grains in a segment are valid. As data become 
invalid, controllers decrement the number of valid grains. When no valid
grains remain, the segment is freed and returns to the \SCM.
Internal fragmentation can lead to inability to allocate new segments, even if
there is enough free space. As is common in log-structured systems, we
free space with a background \GC process.

\subsection{Garbage Collection (GC)}
\label{ssec:gc}

\GC is responsible for relocating fragmented data to provide free segments. The
\SCM executes the \GC algorithm that selects the best segments to relocate; \GC
operates across all devices but independently for each storage type. When a
segment is selected, \GC (up)calls the owning controller to relocate the valid
data of this segment to a new one (\ding{184}). The \GC is not aware of which
grains are valid and which are not, nor the segment geometry in terms of page
size, metadata, etc. This is left to the controller. Once the controller
relocates data, it frees the corresponding grains, and, eventually, segments are
freed and returned to their corresponding free queues (\ding{185}).

%

\GC maintains a number of spare segments for relocation, because
otherwise it will not be able to provide free segments for allocation.  As
with most TLs, \Sys over-provisions storage: it exposes only part of the device total
capacity to the user, and uses the rest for \GC.

Initially, all the device segments are free, and \Sys redirects user writes to
free segments. 
When free segments run out, however,
\Sys \GC needs to perform relocations to clean up segments. Relocations cause I/O amplification and the
underlying devices serve both user and relocation traffic. \Sys uses two
(configurable) watermarks: a low (high) watermark to start (stop) \GC.
For SMR sequential segments of host-managed drives, we reset the zone write
pointers of a segment before placing it in the allocation queue.
\Sys uses a generalized variant of the greedy~\cite{Chang2004tecs} and circular
buffer (CB)~\cite{rosenblum92:logfs} algorithms, which augments a greedy policy
with the aging factor of the CB. This aging factor improves the performance of
the algorithm under a skewed write workload without hindering its performance
under random writes.

\subsection{LSA controller}
\label{ssec:lsa-controller}

\Sys supports multiple front-ends, but in this paper we focus on the Linux kernel
front-end where each controller exposes a block device. These controllers
maintain a mapping between user-visible logical block addresses (LBAs), and
backend physical block addresses (PBAs). We refer to them as Log
Structured Array (LSA)\cite{menon95:lsa} controllers.
LSA controllers map LBAs to PBAs, with a flat array of 32 (default) or 64 bits
for each entry (compile-time parameter). Larger blocks (or pages) require less
space for the table, but lead to I/O amplification for writes smaller than the
page size (e.g., read-modify-write operations for writes). For SSDs, the
default page size is 4KiB, allowing us to address 16TiB (64ZiB for 64 bit
entries) storage; for SMR drives, the default page size is 64KiB. Note that the
page size has to be a multiple of the \SCM grain size, in order to maintain
interoperability with the \GC and allocators. The mapping table is maintained
in-memory, with an overhead of 4B per LBA (e.g., 1GiB DRAM per 1TiB
of storage for 4KiB pages, 512MiB for 8KiB, etc.). A back-pointer table of
PBA-to-LBA mappings is maintained per segment for \GC and restore operations,
and it is either always in-memory or is constructed on-demand by scanning the
LBA-to-PBA table, based on a run-time configuration parameter.

Accesses and updates to the mapping table are done in a thread-safe
lock-free manner using compare-and-swap.  A read operation will
typically read the table, perform a read I/O operation to fetch the necessary
data, and return them to the user. A write operation will allocate new space,
perform a write I/O operation to write user data to this space, and update the
table entry. A relocation operation on a PBA will read the PBA-to-LBA
back-pointer, check that the LBA stills maps to the PBA in question, read the
data, allocate new space, write the valid data to a new location, and update
the table only if the LBA still maps to the relocated PBA.

For sequential segments of host-managed SMRs we \emph{force} the allocated pages
to be written sequentially to the drive, to avoid drive errors. We do so via
a thread that ensures that all writes to these segments happen in-order.  This
is not required for other storage types (e.g., SSDs), and we do not use the
I/O thread for them.


\subsection{Persisting metadata}
%
The LSA controller we described so far maintains the LBA-to-PBA
mapping in memory and dumps it to storage upon shutdown.
%
%
To protect against crashes, controllers log updates to the
mapping table. Under this scheme, a segment contains two types of pages: pages
written by the user, and metadata pages that contain mapping updates. In the
absence of flush operations (e.g., {\ttfamily fsync}), one metadata page is
written for every $m$ data pages ($m$ is configurable at run-time for each
controller). In case of a flush, a metadata page is written immediately.
Therefore, \Sys provides the same semantics as traditional block devices
that use internal write buffers.
The metadata flush is handled differently for different storage \emph{types}.
For SMR storage, we pad segments so we adhere to the sequential pattern.
For Flash, we update the metadata page in-place; although this might break the
strict sequentiality of writes at the SSD level, flush operations are rare, and
did not noticeably affect performance in any of our experiments.
\Sys also maintains a configuration metadata page at each device, and a
configuration metadata page per segment.  The metadata overhead depends on the
value of the $m$, on the segment size, and on the grain size. For 1GiB segments,
$m=512$ (default value), and the smallest possible grain size (4KiB), it amounts
to ~0.2\% of the total capacity.

Upon initialization, we first check whether \Sys was cleanly stopped using
checksums and unique session identifiers written during the LBA-to-PBA
dumps. If a clean shutdown is detected, the mapping of each controller is
restored.  Otherwise, \Sys scans for metadata pages across all valid segments
belonging to the controller, and restores LBA-to-PBA mappings based on
back-pointers and timestamps. The \SCM coordinates the restore process: it
iterates over segments in parallel and upcalls owning controllers to restore
their mapping.

\subsection{Implementation notes}

The core of \Sys is implemented as a library that can run
in kernel- or user-space. Different front-ends provide different
interfaces (e.g., a block device, or a key-value store) to the user. The Linux
kernel block device interface is implemented on top of the device-mapper (DM)
framework.  \Sys controllers are exposed as DM block devices. Any I/O to these
devices is intercepted by the DM and forwarded to the \Sys kernel module, which
in turn remaps the I/O appropriately and forwards it to the underlying physical
devices. Devices can be created, managed and destroyed, using the SALSA user interface tool (UI).

%
%

\section{Adapting to application workloads}
\label{sec:adaptation}

Host TLs can be adapted to different application workloads, which we fully
embrace in \Sys.
At a first level, \Sys offers a large number of parameters for run-time
configuration. Controllers parameters include:
page size, number of streams for user/\GC writes, metadata stripe size,
sets to specify storage types each controller can use, etc.
Furthermore, \Sys includes multiple controller implementations, each with their
own specific parameters. There are also global parameters: grain size, \GC
implementation (each with its own parameters), \GC watermarks, etc.  Users are
not expected to understand these details: the UI provides sane
default values.  In practice, we have found this
rich set of options extremely useful.
Moreover, \Sys can be extended to adapt to specific workloads and meet
different application requirements by implementing different controllers. For
example, an RDMA interface to NVM storage has been implemented as a \Sys
controller in FlashNet~\cite{Trivedi17flashnet}. Next, we discuss two controller designs
that we developed to address application-specific workloads.
%

\subsection{Dual mapping controller}
\label{ssec:sys:object-store}
\label{ssec:sys:dual-mapping}

Our use-case is running an object store service on SMR drives.
Because object stores perform fault management and load distribution,
we run one \Sys instance per drive and let the upper layer balance load and deal
with faulty drives.
For object stores, 128KiB is considered a small object
size\cite{zheng13:cosbench}. Therefore, we can set the page size to 64KiB,
leading to an overhead of  64MiB RAM per 1TiB of storage, making \Sys feasible
even for servers that can contain close to a hundred drives\cite{onestor}.



During an initial evaluation, we observed a number of read-modify-write
operations that degraded performance. We found two sources for this: writes that
are not aligned to 64KiB, and filesystem metadata that are smaller than
64KiB. Even though the number of these operations is relatively small, they lead
to a noticeable performance hit.
%
%
We can avoid read-modify-write operations with a controller that supports a
small number of sub-64KiB mappings, while using 64K pages for everything else.
To that end, we develop a \emph{dual mapping controller} that maintains two
mappings: a sparse mapping for 4KiB pages and a full mapping for 64KiB. A read
operation checks whether the pages exist in the sparse mapping \emph{first} and
if they do not, checks the full mapping. A write operation will use the full
mapping for 64KiB pages, and the sparse mapping for smaller pages. If the sparse
mapping does not contain a page during a write and has no free locations, we
perform a read-modify-operation and update the full mapping.

\subsection{Hybrid controller}
\label{ssec:sys:vidserver}
\label{ssec:sys:hybrid}

Hybrid systems that make use of multiple storage types allow tradeoffs that are
not possible otherwise, offering great opportunities for maximizing the
system's utility\cite{fbtimeline,ripq:fast15}.
%
As storage diversity increases, we expect the importance of hybrid systems to
rise. For example, vendors offer hybrid drives (Solid State
Hybrid Drives -- SSHDs) that combine a disk and NAND Flash within a single
drive\cite{seagete:sshd,wdblack2}. These drives, however, have hard-coded
policies and cannot be re-purposed.

Multi-device host TLs enable building better hybrid systems. In contrast to a
device TL, a host TL can support multiple devices from different vendors.
Moreover, the indirection and mechanisms employed by a TL like \Sys can be
reused, enabling transparent data relocation between different media.
Finally, as we argued throughout this paper, a host implementation offers
additional flexibility, as well as co-design potential\cite{saxena12:flashtier},
compared to a drive implementation.

We consider a video service for user-generated content (e.g., YouTube).
Because user-generated videos are typically short\cite{cheng13:youtube}, and
will get striped across a large number of disks, reading them from a disk will
result in reduced throughput due to seeks. Because most views are performed on
a relatively small subset of the stored videos, there is an opportunity to
optimize the read throughput by moving them into a faster medium. If the read
working set does not fit to DRAM, moving data to an SSD is
the best solution. The next section presents a \Sys controller
implementing this functionality.

For our hybrid controller, we configure two allocation streams: one for fast
storage (Flash) and one for slow storage (disks). User and \GC relocation writes
always allocate from the slow storage stream, while the fast storage stream is
used for relocating ``hot'' pages that are frequently accessed.
To determine the ``hot'' pages we maintain a data structure with
a ``temperature'' value of each (logical) page. We use an array
of 64-bit values sized at the number of logical pages divided by 256
(configurable) to reduce memory overhead. Because most files are stored
sequentially on the block device, we map consecutive pages to the same
temperature.

At each read, we increase the temperature by one. Because we use 64-bit values,
overflowing is not an issue. If a page is discarded, we set the temperature to
zero. We also periodically (once a day) halve the temperature of all values.
%
When we increase the temperature, we check the new value against a configurable
threshold. If the threshold is reached and the page is not already
located into Flash storage, we schedule a relocation. The relocation happens
asynchronously on a different (kernel) thread to avoid inducing overhead to the
read operation. If at any point something goes wrong (e.g., there are no
available Flash physical pages, the mapping or temperature changed in the
meantime) the operation is aborted.

%% file: evaluation.tex
\begin{figure*}[h!tpb]
   \subfloat[100\% writes]{%
       \includegraphics[width=.48\textwidth]{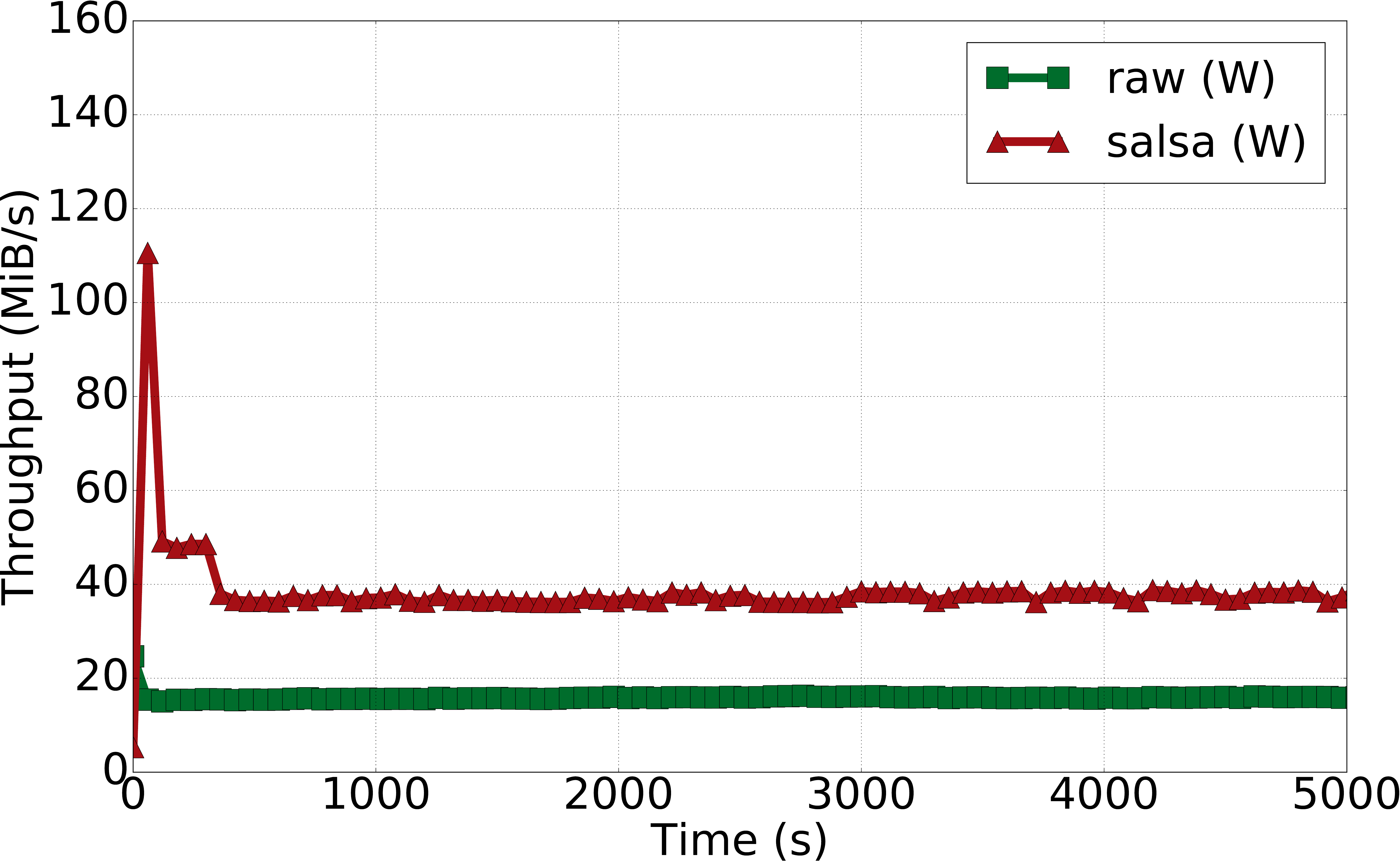}
       \label{fig:eval:ubench:ssd:wr}
   }
   \hfill
   \subfloat[20\% writes - 80\% reads] {%
       \includegraphics[width=.48\textwidth]{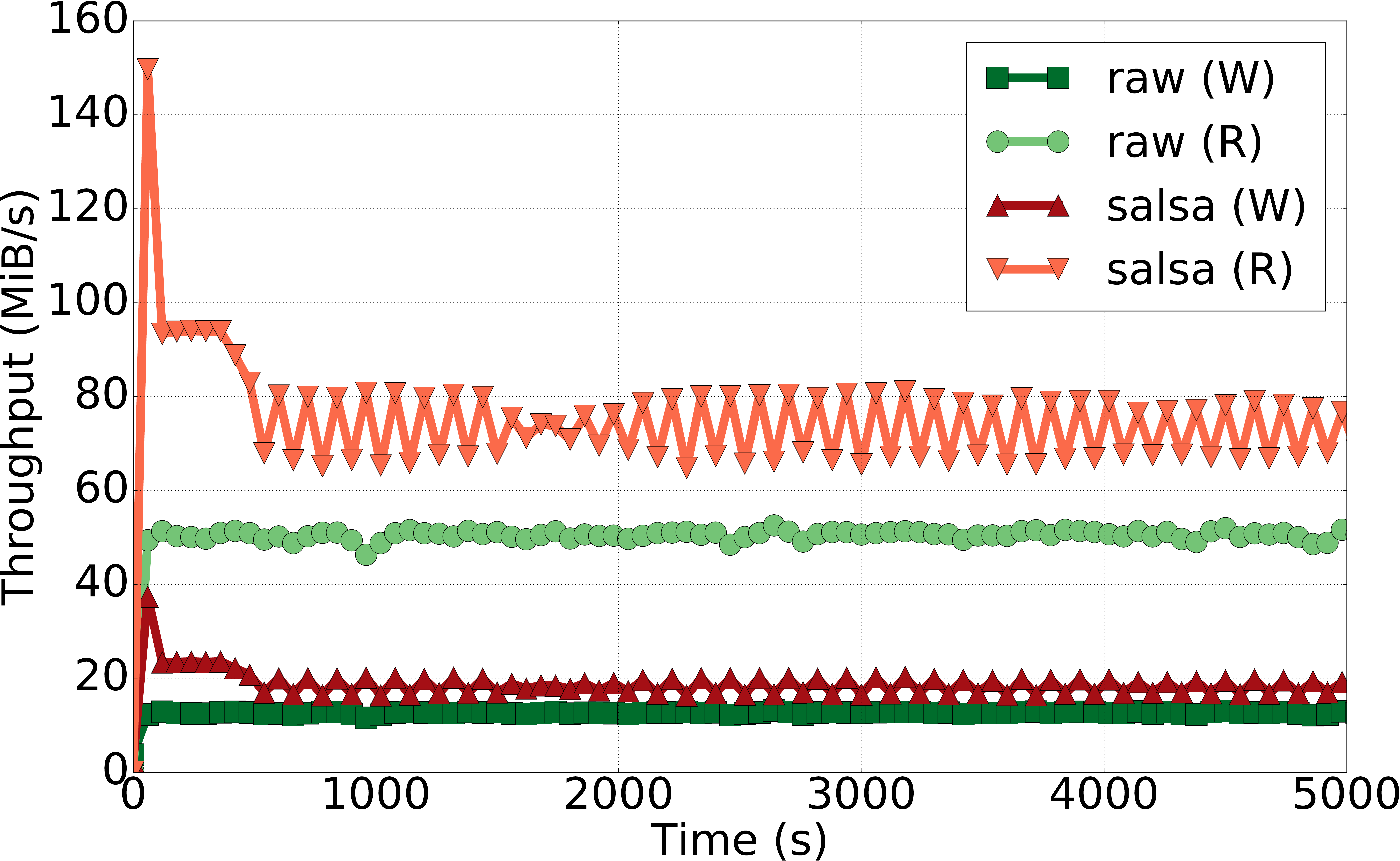}
       \label{fig:eval:ubench:ssd:rd}
   }
   \caption{4KiB uniform random workload on an SSD}
   \label{fig:eval:ubench:ssd}
\end{figure*}

\section{Evaluation}
\label{sec:evaluation}

We start our evaluation (\Sec{ssec:eval:ubench}) discussing the performance and
durability benefits of \Sys using a random workload.
Subsequently, we show how \Sys features can benefit real-world applications.
In \Sec{ssec:eval:mysql}, we evaluate the benefits of \Sys's single- and
multi-device optimizations using  MySQL containers on SSDs.
In \Sec{eval:object-store}, we evaluate the dual-mapping controller
(\Sec{ssec:sys:dual-mapping}) using Swift.
We use both MySQL and Swift to evaluate the benefits of supporting multiple
application policies in \Sec{sec:eval:dual-controller}.
Finally, in \Sec{sec:eval:vidserver}, we evaluate the hybrid controller
(\Sec{ssec:sys:hybrid}) for a user-generated video service.

SSD experiments (\Sec{ssec:ubench:ssd}, \Sec{ssec:eval:mysql}, \Sec{sec:eval:dual-controller}) are performed on
a 16 core dual-node x86-64 server with 128GiB RAM running RHEL 7.2 with a 3.10
Linux kernel, using a widely-used off-the-shelf 2.5'' 1TB SATA NAND
Flash SSD.
The SMR experiments (\Sec{ssec:ubench:smr}, \Sec{eval:object-store},
\Sec{sec:eval:vidserver}) are performed on a 4 core x86-64 server with 20GiB
RAM running RHEL 6.6 with a 4.1 kernel, with a drive-managed 8TB SMR drive.

\subsection{Microbenchmarks}
\label{ssec:eval:ubench}

To evaluate the benefits of placing the TL on the host, we compare the
performance and endurance of \Sys against raw SSD and SMR drives under a
sustained random workload.
We use this workload because random writes are
highly problematic for SSDs and SMRs for two reasons.
First, \GC runs
concurrently with user operations and causes maximum disruption.
Contrarily, in a bursty workload, \GC would have time to collect between
bursts. Second, random writes across the whole device maximize write
amplification.
We use a microbenchmark that applies uniformly random
read and writes directly (using \pr{O\_DIRECT}) to the device.
We measure device throughput using \pr{iostat}\cite{iostat}.

\subsubsection{SSDs}
\label{ssec:ubench:ssd}

\newcolumntype{?}{!{\vrule width 2.5pt}}
\begin{table}[b]
\centering
\small
\begin{tabular}{|l|r|rr|}\hline
                        & write-only    & \multicolumn{2}{c|}{read-mostly} \\\hline
        throughput      & W:100\%       &  R:80\%        &  W:20\%         \\\hline
\rowcolor[gray]{.9}
         raw            & 15.9 $\pm$ 0.2  &  50.6 $\pm$  0.9 &  12.6 $\pm$  0.2  \\\hline
       salsa            & 37.7 $\pm$ 0.9  &  72.5 $\pm$  6.2 &  18.1 $\pm$  1.5  \\\hline
\end{tabular}
\caption{Average throughput (MiB/s) and standard deviation for two random
workloads on an SSD: 100\% writes and 80\%/20\% reads/writes. Block size is 4KiB.}
    \label{tab:eval:ubench:ssd}
\end{table}

We low-level format the drive before our experiments.
We overprovision \Sys with 20\% of the SSD's capacity. We facilitate a fair
comparison by performing all measurements on the raw device on a partition with
equal size to the \Sys device. That is, we reserve 20\% space on the raw device
which is never used after low-level formatting the drive. The 20\%
overprovision was chosen to offer a good compromise between \GC overhead and
capacity utilization~\cite{radu-freq}.
%
To measure stable state, we precondition the device (both for raw and \Sys)
by writing all its capacity once in a sequential pattern,
and once in a uniformly random pattern. Subsequent random writes use
\emph{different} patterns.

We consider two workloads: \emph{write-only} (100\%~writes) and
\emph{read-mostly} (80\%~reads, 20\%~writes), both with 4KiB blocks and queue
depth (QD) of 32.  The benefits of \Sys in a \emph{read-mostly}
workload are smaller because read operations do not directly benefit from \Sys
and write amplification due to \GC having a smaller impact when writes are
infrequent.

Stable state throughput over time is shown in \Fig{fig:eval:ubench:ssd}, and
the average throughput is reported in \Tab{tab:eval:ubench:ssd}. \Sys
achieves $2.37 \times$ better average throughput than the raw device for a
write-only workload.
For a read-mostly
workload, \Sys improves both read and write throughput by $1.43 \times$.  We
attribute the worse read throughput of the raw device to obstruction caused by
the drive \GC that stalls reads.
Moreover, we have extensively experimented with more than 20 commodity SSDs.
Among those, using 20\% overprovisioning, \Sys improves throughput on a
sustained random write workload by a factor of $1.5 \times$-$3 \times$.

Next, we compare endurance when using \Sys
against using the raw drive.
We measure wear via a SMART attribute that, according to
the device manufacturer, increases linearly with the wear (Program/Erase cycles)
of the Flash cells. We low-level format the drive and fill it up once
sequentially.  Subsequently, we perform 10 full device random writes with 4KiB.
We record the wear of the device after each full device write (11 data
points including the initial sequential write). We repeat the experiment 6 times
alternating between runs on the raw device and runs on \Sys. As before,
experiments on the raw device were performed on a partition equal to the \Sys
device size, so a full device write amounts to the same amount of data in both
cases.

\begin{figure}
    \centering
    \includegraphics[width=\linewidth]{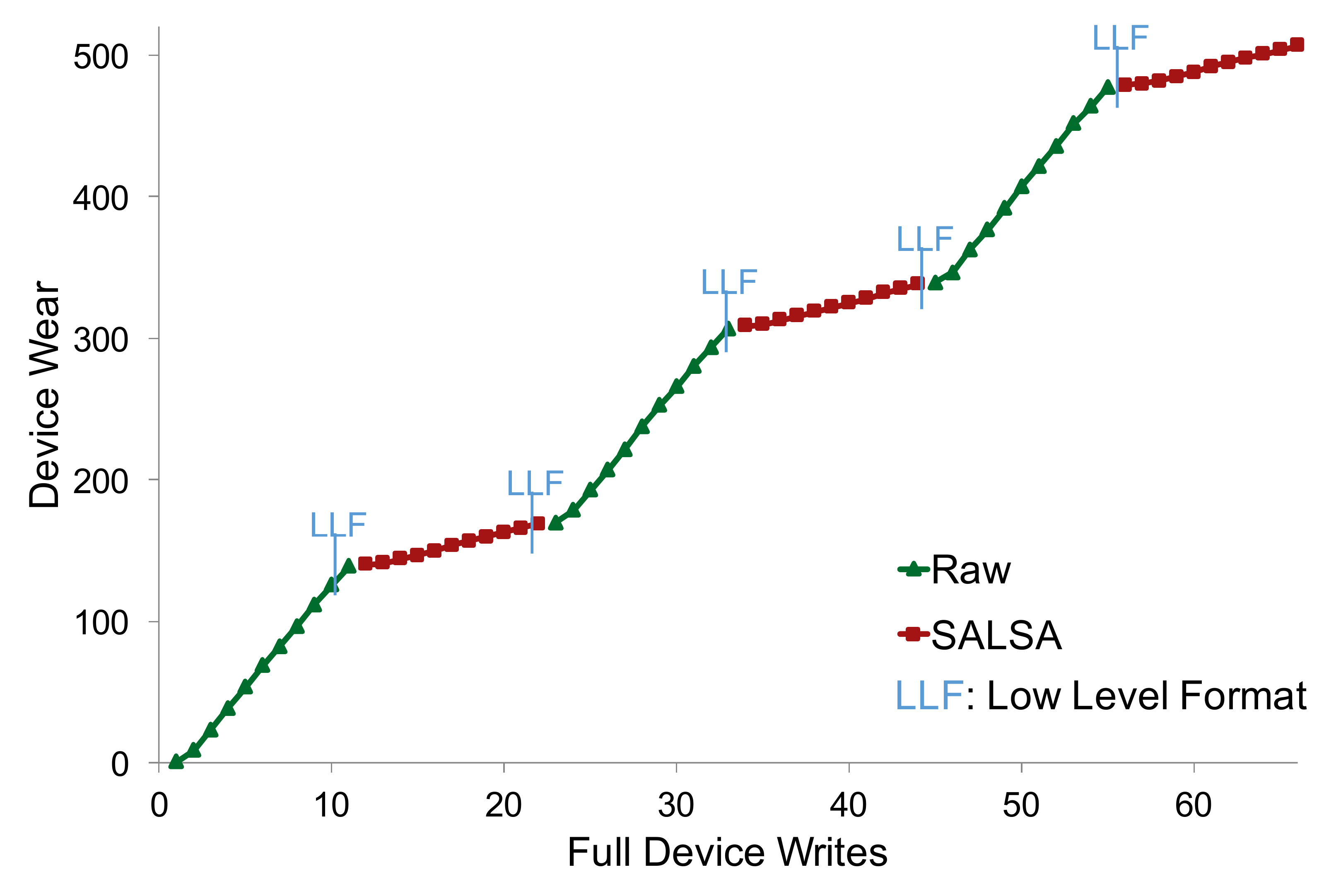}
    \caption{SSD wear with and without \Sys.}
    \label{fig:eval:ubench:endurance}
\end{figure}

Results are shown in \Fig{fig:eval:ubench:endurance}. Overall, the same
workload incurs $4.6 \times$ less wear to the device when running on \Sys
compared to the raw device. In this experiment, we measured a write
amplification of 2.5 on average for \Sys (which is very close to the
theoretically expected 2.7 for random writes and chosen
overprovision~\cite{radu-freq}), which implies that the internal drive write
amplification was $11 \times$ less compared to the raw device experiment; \Sys
wrote $2.5 \times$ the user data and still induced $4.6 \times$ less total
device writes compared to the raw device, suggesting that the total device
writes for the raw device was $2.5 \times 4.6 \approx 11 \times$ the user
data. Note that the internal drive write amplification typically includes
metadata (and possibly data) caching on top of GC traffic; in fact, since the
\GC traffic should be similar between the two experiments for random writes, we
attribute most of the extra amplification to this cache traffic. Results were
repeatable over multiple executions of the experiment, and other commodity SSDs
we examined behaved similarly.

\subsubsection{SMRs}
\label{ssec:ubench:smr}

We now turn to SMR drives, comparing the performance of \Sys against the raw
device using 64KiB uniform random writes with QD1 across the
whole device.

We use \Sys with all SMR variants (drive-managed, host-aware, and
host-managed) across multiple vendors. Here, we present results for a drive-managed
SMR, because we can directly compare against the drive's TL by applying the
same workload on the raw device.\footnote{The host-aware SMR drives that we
tested were vendor samples, and therefore might not be representative of the
final products. For this reason we opted to present results for widely-available
drive-managed SMRs.}
A drive-managed SMR, however, limits \Sys because it does not
expose drive information (e.g., zones)
and cannot be directly controlled (e.g., does not allow resetting write
pointers). Instead, similarly to SSDs, \Sys writes sequentially to minimize
the drive's TL interference.
We overprovision \Sys by 10\%, and low-level format the drive before the
experiments.
We select this value with a steady-state random workload in mind; for other
workloads (e.g., read-mostly or sequential) smaller values might offer a better
tradeoff.

\begin{figure}[tpb]
    \centering
    \includegraphics[width=.85\linewidth]{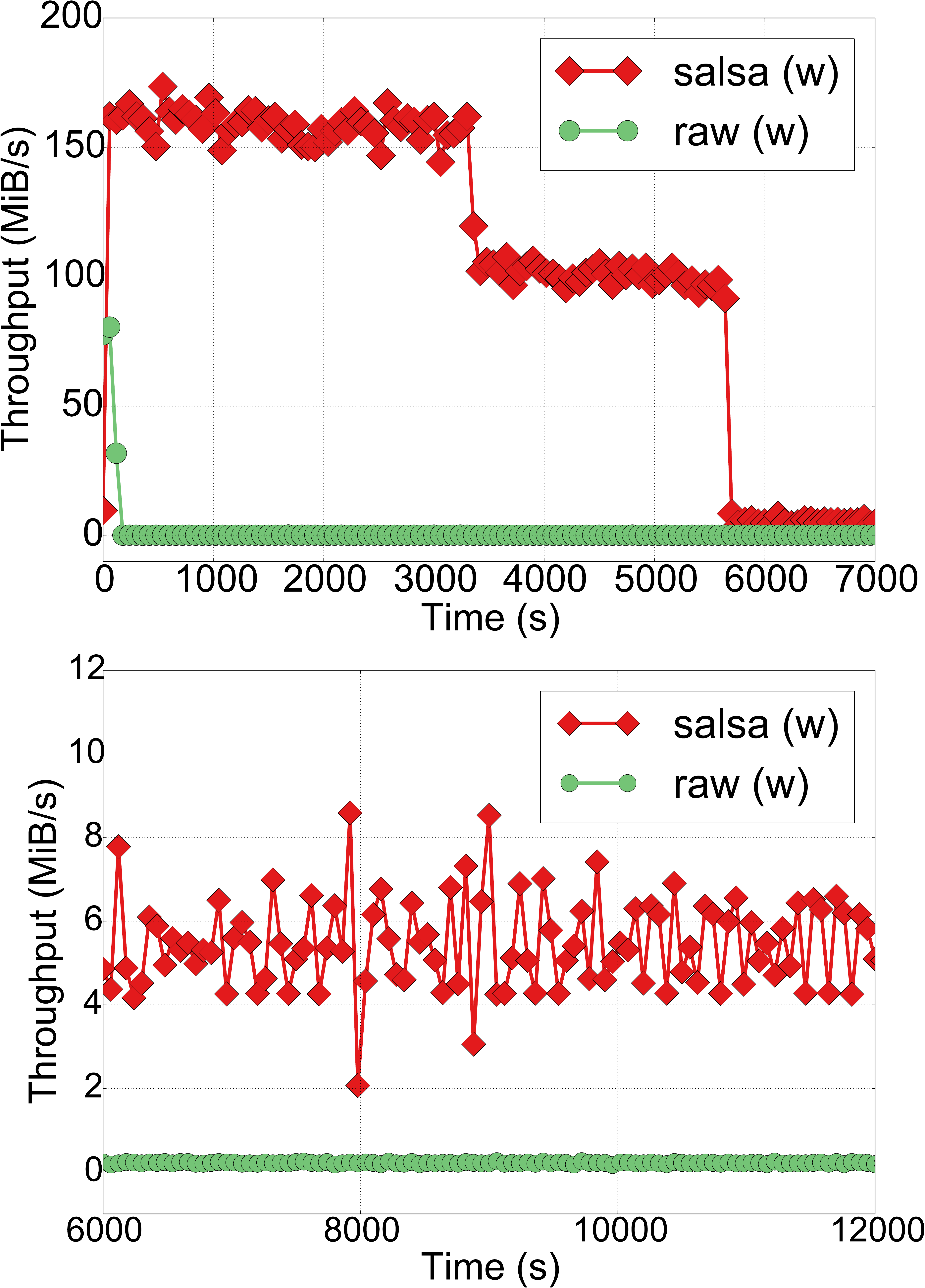}
    \caption{64KiB random writes on a host-managed SMR with and without \Sys. The
    raw results are after the first write on the device, while the \Sys results
    are after the whole device was randomly written once. The top plot shows
    the 0-7~Ksecs area, while the bottom focuses on the
    6-12~Ksecs area.}
    \label{fig:eval:ubench:smr}
\end{figure}

The results are shown in \Fig{fig:eval:ubench:smr}.
The raw device throughput starts close to 80MiB/s but drops to
200~KiB/sec after about 5 minutes, which renders
the device effectively unusable for many applications. We attribute the drop in
performance to the persistent cache of the drive, as identified by prior
work~\cite{skylight:fast15, aghayev:fast17}: after the persistent cache is
filled ($\sim1.4GiB$ of random 64KiB writes~\cite{skylight:fast15}), then
the drive starts its cleanup process, which entails read-modify-writes on MiBs
of data.

Contrarily, \Sys's performance does not degrade that quickly.
Hence, to facilitate an easier comparison \Fig{fig:eval:ubench:smr} presents
\Sys throughput results \emph{after} a full device (random) write.
We observe three phases in \Sys performance.
During the first and second phase, no \GC is performed.
Initially, caching on the drive
allows an initial throughput of roughly 160MiB/s, which drops to 100MiB/s after
about 3K seconds. This designates the \Sys performance for bursts up to an
amount of data equal to the difference between the high and low \GC watermarks.
In the third phase, \GC starts and the throughput of the \Sys device becomes
roughly 5~MiB/s, $25 \times$ better than the throughput of the raw drive.


\subsection{Containerized MySQL on SSDs}
\label{ssec:eval:mysql}

In this section, we evaluate the effect of \Sys's single- and multi-device
optimizations on the performance of a real-world database.
Specifically, we deploy multiple MySQL Docker containers on commodity SSDs in a
single- and a multi-device (RAID-5) setup, and execute an OLTP workload
generated by sysbench~\cite{sysbench}.

We evaluate 5 container storage configurations: three with
1 SSD (raw device, F2FS~\cite{f2fs:2015}, and \Sys), and two with
4 SSDs using RAID-5
(Linux MD\cite{md} and \Sys equivalent).
We use the same hardware and
setup (formatting, partitioning, preconditioning) as in~\ref{ssec:ubench:ssd}.
For F2FS, we also allocate the same over-provisioning as the other deployments: 20\% using the \pr{-o 20} option when creating the filesystem with \pr{mkfs.f2fs}.
We only use F2FS in the single device deployment, since it did not provide a native RAID-5 equivalent multi-device deployment option.
The only difference across our experiments is the device we use (a raw
device partition, a \Sys device, or an MD device). In this device we create one
(logical) volume per MySQL instance to store database data.
%
We use the Ubuntu 14.04 image provided by Docker, adding the necessary packages
for our experiment.
We deploy four containers with one multi-threaded MySQL server per container.
Each server uses a 160GiB database image which we place on the corresponding
volume.  On each container, we run 4 sysbench threads to maximize IO
throughput.
We use the default LSA controller (\Sec{ssec:lsa-controller}) for \Sys.

\begin{figure*}
    \subfloat[1 SSD: raw device, F2FS, and \Sys]{%
        \includegraphics[width=.48\textwidth]{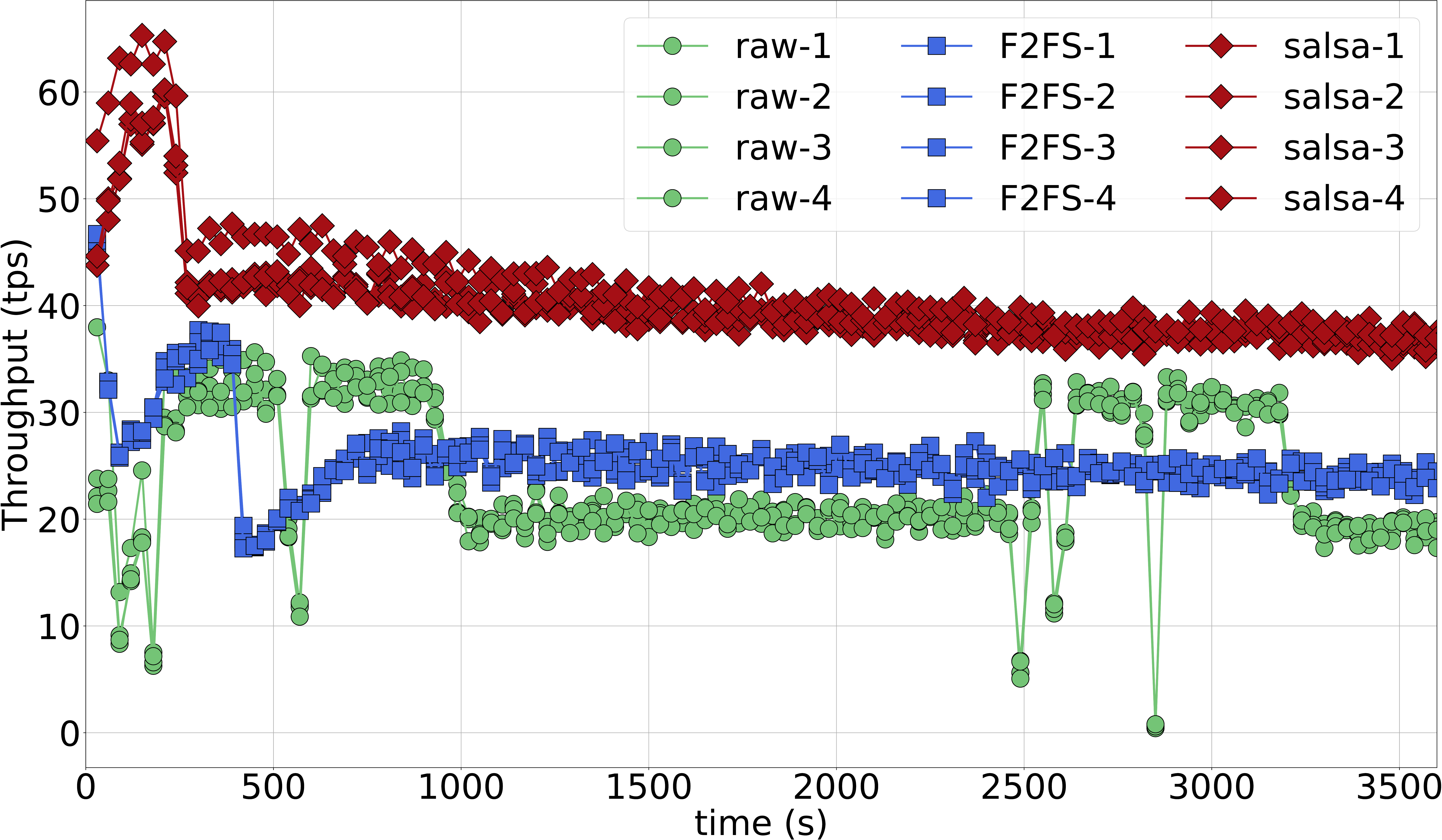}
        \label{fig:eval:sysbench-1ssd:tps}
    }
    \hfill
%
    \subfloat[4 SSDs: Linux md software-RAID 5 and \Sys RAID-5 equivalent]{%
        \includegraphics[width=.48\textwidth]{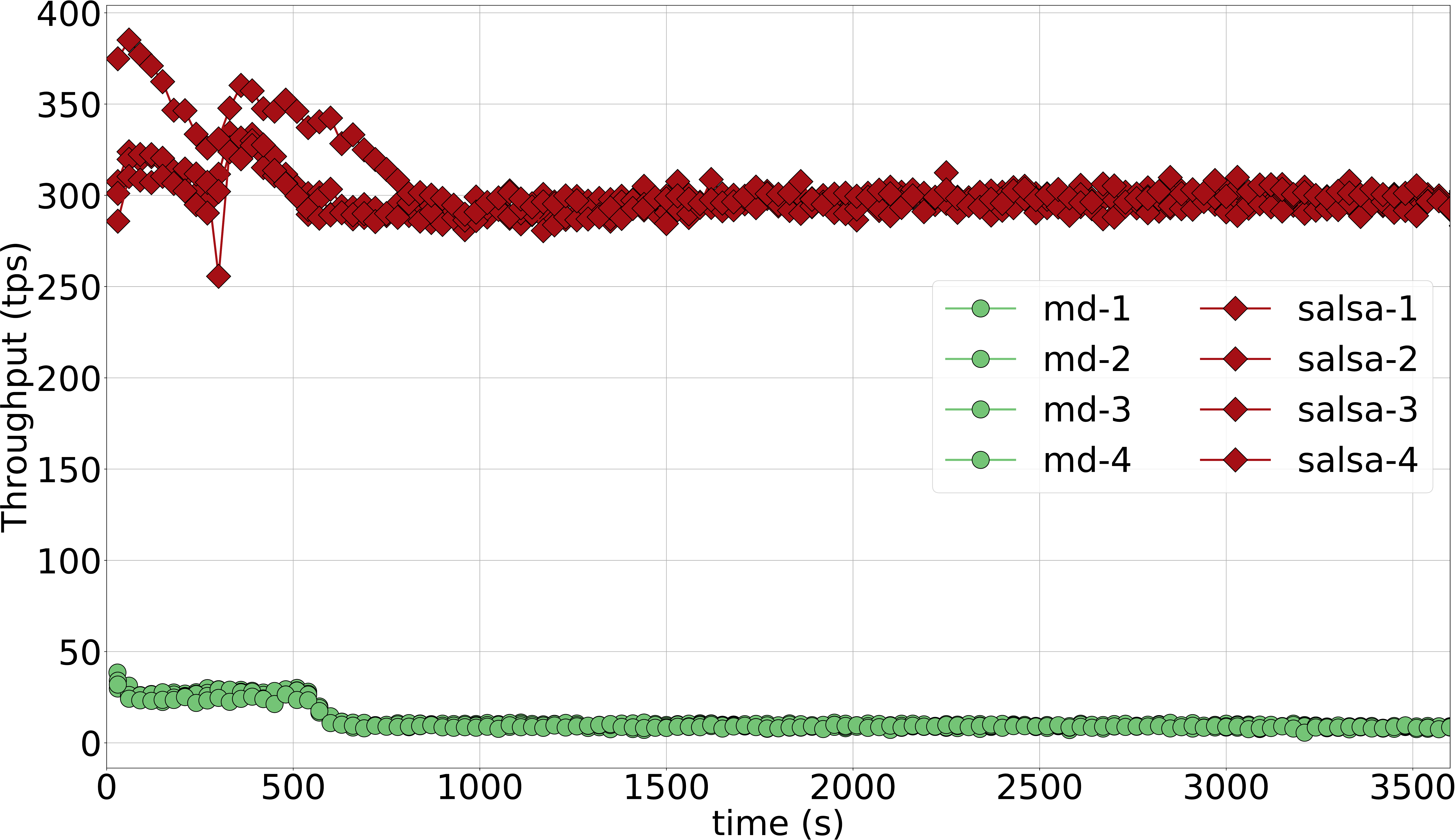}
        \label{fig:eval:sysbench-raid:tps}
    }
    \caption{Throughput of sysbench during execution}
    \label{fig:eval:sysbench}
\end{figure*}

\begin{table}[h]
    \footnotesize
    \centering
    \subfloat[1 SSD]{%
      \setlength\tabcolsep{4pt} 
        \begin{tabular}{|ccc|ccc|ccc|}\hline
        \multicolumn{3}{|c|}{raw} & \multicolumn{3}{c|}{F2FS} & \multicolumn{3}{c|}{\Sys} \\\hline
                tps &      avg &      95\% &   tps &      avg &     95\% &   tps &      avg &     95\% \\
\rowcolor[gray]{.9}
               22.2 &  180ms &   651ms &  25.6 &  157ms &  599ms &  37.4 &  107ms &  266ms \\
               21.3 &  188ms &   655ms &  25.6 &  156ms &  599ms &  37.6 &  106ms &  264ms \\
\rowcolor[gray]{.9}
               21.2 &  188ms &   656ms &  25.5 &  157ms &  596ms &  37.7 &  106ms &  264ms \\
               21.2 &  188ms &   654ms &  25.6 &  157ms &  603ms &  39.1 &  102ms &  258ms \\
        \hline
        \end{tabular}
        \label{tab:sysbench-1ssd}
    }
%

    \subfloat[3+1 SSDs RAID-5: Linux MD and \Sys]{%
        \begin{tabular}{|rrr|rrr|}\hline
        \multicolumn{3}{|c|}{Linux MD} & \multicolumn{3}{c|}{\Sys} \\\hline
              tps &   avg & 95\% &   tps &    avg &    95\% \\
\rowcolor[gray]{.9}
              8.1 &  2.0s & 5.3s & 287.2 & 55.7ms &  99.5ms \\
              8.1 &  2.0s & 5.3s & 290.5 & 55.1ms &  98.4ms \\
\rowcolor[gray]{.9}
              8.3 &  1.9s & 5.2s & 286.5 & 55.9ms &  99.9ms \\
              7.8 &  2.1s & 5.6s & 291.1 & 55.0ms &  98.2ms \\
        \hline
        \end{tabular}
        \label{tab:sysbench-raid}
    }
\caption{Sysbench results for each MySQL instance: throughput
in transactions per second (\emph{tps}), average (\emph{avg}) and
95th percentile (\emph{95\%)} response times.}
\label{tab:sysbench}
\end{table}

Results for one SSD, as reported by each sysbench instance are shown in
\Tab{tab:sysbench-1ssd}. \Fig{fig:eval:sysbench-1ssd:tps} depicts sysbench
throughput over time for each instance.  \Sys improves throughput by $1.68
\times$, and the average latency by $1.69 \times$ compared to raw device,
illustrating the benefits of implementing a TL on the host, instead of the
device where resources are limited.  Also, \Sys provides an improved throughput
by $1.47 \times$ compared to F2FS, at a reduced tail latency (95\% percentile)
of $2.45 \times$. We attribute the improvement against F2FS mainly to two
reasons: (i) F2FS uses a 2MiB segment size which is not optimal for modern
commodity SSDs~\ref{ssec:bg:ftl-limitations}, compared to segments at the GiB
level for \Sys, and (ii) F2FS updates its metadata in separate write-logs and
at eventually in-place~\cite{f2fs:2015} which further reduce the effective
sequential I/O size as received at the drive TL level; large, uninterrupted
sequential overwrites are essential to achieve the ideal write performance of
non-enterprise grade SSDs~\cite{andromeda17nvmw}.

\Fig{fig:eval:sysbench-raid:tps} and \Tab{tab:sysbench-raid} show results for
four SSDs in a RAID-5 configuration, using Linux MD and \Sys.  \Sys increases
throughput by $35.4 \times$ and improves the average response time by $36.8
\times$.
These results showcase the significant benefits of a TL that is multi-device
aware.
While \Sys can guarantee full-stripe writes with a small
persistent buffer, in-place update approaches such as Linux MD cannot, because
that would require a
buffer with size comparable to device capacity. Hence, in-place updates in
Linux MD trigger read-modify-write operations that lead to response times in the
order of seconds, rendering this setup unsuitable for many applications.


%
We also note that the performance difference between \Sys for one device and
RAID-5 is due to the lower \GC pressure in the latter case, since the RAID-5
configuration has 3 times the capacity of the single device configuration
while the working set size does not change across the two tests.  Contrarily,
the Linux RAID-5 implementation has lower throughput than the single device,
due to the parity updates and read-modify-write operations, which also slow
down dependent reads.

Finally, the CPU overhead is negligible. In the RAID-5 configuration, we
measured an overhead of less than 6\% in normalized CPU utilization (CPU
utilization / TPS) compared to the raw Linux MD configuration.

\subsection{Object store using SMR drives}
\label{eval:object-store}

A host TL enables workload-specific optimizations. We evaluate
the benefits of such an approach by running an object store on SMR disks,
comparing the \Sys dual-mapping controller (\Sec{ssec:sys:object-store}) against
the raw device.

We use Openstack Swift, a popular open-source
eventually-consistent object store\cite{openstack-swift}. Swift is written
in Python and includes multiple services. For our evaluation we
focus on the object server\cite{openstack-swift-arch},  the component that
stores, retrieves, and deletes objects on local devices. Objects are stored
as files on the filesystem, while object metadata are stored in the file's
extended attributes.
For both experiments, we use an XFS filesystem configured per
the Swift documentation\cite{openstack-sw-conf} on the same SMR drive
as \Sec{ssec:ubench:smr}.
To isolate storage performance, we
wrote a Python program that issues requests directly to the object server.
Swift uses ``green threads'', i.e., collaborative tasks, to enable concurrency.
We do the same, using a 32 green thread pool for having
multiple requests in flight.

We initialize the object store via \pr{PUT} operations, so that the total data
size is 64GiB. We subsequently update (\pr{UPDATE}), and finally retrieve
(\pr{GET}) all the objects. The last two operations are performed in different
random orders. We clear the filesystem and block caches before starting each
series of operations.

\newcommand{\xfname}[1]{results/obj_bench/obj_bench_logs.theseus.zurich.ibm.com.20161010.184039/r0-#1}
\begin{figure}
    \centering
    \includegraphics[width=\columnwidth]{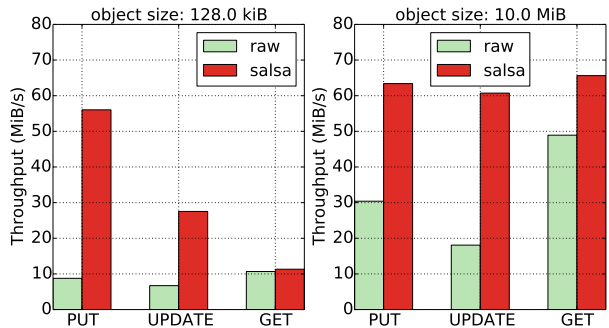}
    \caption{Swift storage server throughput for different operations, comparing the raw device
    and \Sys.}
    \label{fig:swift-throughput}
\end{figure}

\begin{figure*}[t]
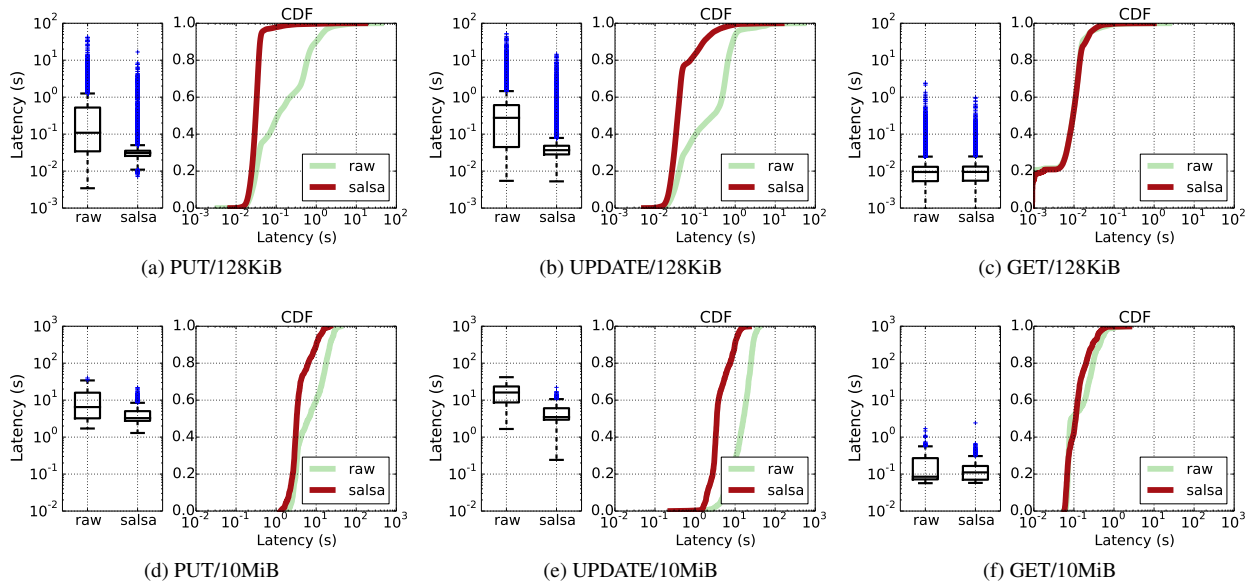

    \subfloat[PUT/128KiB]{%
        \includegraphics[width=.32\textwidth]{\xfname{s131072-PUT-crop}}
    }
    \hfill
    \subfloat[UPDATE/128KiB] {%
        \includegraphics[width=.32\textwidth]{\xfname{s131072-UPDATE-crop}}
    }
    \hfill
    \subfloat[GET/128KiB] {%
        \includegraphics[width=.32\textwidth]{\xfname{s131072-GET2-crop}}
    }

    \subfloat[PUT/10MiB]{%
        \includegraphics[width=.32\textwidth]{\xfname{s10485760-PUT-crop}}
    }
    \hfill
    \subfloat[UPDATE/10MiB] {%
        \includegraphics[width=.32\textwidth]{\xfname{s10485760-UPDATE-crop}}
    }
    \hfill
    \subfloat[GET/10MiB] {%
        \includegraphics[width=.32\textwidth]{\xfname{s10485760-GET2-crop}}
    }

    \caption{Swift storage server latency for different operations, comparing the raw device
    and \Sys. The box is placed in the first and third quartiles, the line inside the box is the median, and the whiskers are at 1.5 IQR.}
    \label{fig:eval:object-store-lat}
\end{figure*}

\Fig{fig:swift-throughput} shows the throughput of the raw drive and \Sys for
128KiB and 10MiB objects for each different operation.
(These sizes were found to represent small and large object sizes in the
literature\cite{zheng13:cosbench}.)
For small objects, using the raw device leads to low throughput for both
\pr{PUT}s and \pr{UPDATE}s: 8.8 and 6.7 MiB/s. We attribute the poor
performance to the drive having to write different files, located at different
extents, potentially triggering relocation. \Sys, on the other hand, achieves
higher throughput: 56~MiB/s for \pr{PUT}s ($6.36 \times$) and 27.5 MiB/s
for \pr{UPDATE}s ($4.1 \times$). \pr{UPDATE}s exhibit lower performance for
both systems since the file that represents the object needs to be
modified. \pr{GET} performance is similar for both systems: 10.7 for raw and
11.3 MiB/s for \Sys.
For large objects the behaviour for \pr{PUT}s and \pr{UPDATE}s is similar, but
the difference between the raw device and \Sys is smaller. For \pr{PUT}s \Sys
achieves 63.4 MiB/s, $2 \times$ higher than the raw device (30.4
MiB/s); for \pr{UPDATE}s the respective numbers are 60.7 MiB/s and 18.1 MiB/s, a $3.35 \times$ improvement for \Sys. \Sys results in better throughput for the \pr{GET} operation of
large objects at 65.6 MiB/s, while the raw device is at 48.9 MiB/s. We
believe this is because XFS uses multiple extents for large files. In
\Sys, these extents end up being close together even if they have different
logical addresses, thus minimizing seek time when accessing them.


In addition to throughput, we sample the operation latency every ten operations and
summarize the results in \Fig{fig:eval:object-store-lat}, using
a box plot and a CDF diagram for each operation type. Because latency
has a wide range of values, we use a logarithmic scale.
For small objects, \Sys results in a lower median latency for both \pr{PUT} and
\pr{UPDATE} operations: 30.8ms and 36.8ms. Using the raw device leads to much
higher latencies: 109ms for \pr{PUT} ($3.5 \times$ higher) and 276ms for
\pr{UPDATE} ($7.5 \times$ higher). Both \Sys and raw have a similar median latency
for \pr{GET}: 9.5ms.
%
%
For large objects, \Sys still achieves a significantly lower median latency
that the raw device.  The median latency for a
\pr{PUT} on the raw device is close to $2 \times$ higher than \Sys
(6.5s versus 3.3s), while for \pr{UPDATE}s raw is $4.6 \times$ higher than \Sys
(16.1s versus 3.5s). The raw device achieves an improved latency of 84.8ms for
\pr{GET} compared to \Sys that achieves 111.1ms, but as shown in
\Fig{fig:eval:object-store-lat}, the raw device has a wider spread.

The relation between latency and throughput is different for \pr{GET}s and write
operations (\pr{PUT}s and \pr{UPDATE}s). In small objects, for example,
\pr{GET}s have lower throughput even though they have lower latency. This is
because write operations allow higher concurrency. Swift performs writes by
placing the data into a temporary file, updating metadata, calling \pr{fsync},
and finally moving the file in its proper location using \pr{rename}. The final
steps are offloaded to another thread
and execution continues with the next
request. Only after a number of subsequent requests are serviced, the initial
requests will be allowed to continue execution and complete, even if the
\pr{rename} call was completed before that.  This approach enables
high-throughput but can significantly hurt latency.

\subsection{Multiple TLs for mixed workloads}
\label{sec:eval:dual-controller}

\Sys enables different policies over a storage pool by decoupling
storage policy and space management. Each policy is implemented as a different
controller (TL) that exposes a different device to the user. In this section,
we evaluate the benefits of this approach by deploying two different
controllers on an SSD. Specifically, we run a containerized MySQL database 
on an LSA controller (\Sec{ssec:lsa-controller}) with an 8KiB page size
to match the database page size, and a Swift object storage system on
a dual-mapping controller (\Sec{ssec:sys:object-store}) on the same SSD.
We compare this approach against two others that use traditional
partitions (one for each application): the raw device, and the LSA controller (configured with
the default 4KiB page size). We
run the mixed workload comprising sysbench OLTP and a object store \pr{PUT}
 workload with 128KiB objects for 30 minutes and evaluate the different configurations based
on memory footprint and application performance. We further compare the
observed relocation traffic for \Sys under the two configurations.

\Tab{tab:eval:dual-controller} summarizes the results. For F2FS, we include
results with 1MiB objects since under 128KiB objects its performance was low
(17.46 sysbench tps, and 6.7MiB/s object write throughput), due to stressing
the file creation scalability of the filesystem at hundreds of thousands of
files~\cite{min16atc}, which was not the aim of this evaluation.  Both \Sys
configurations maintain a performance improvement similar to the single-SSD
experiments presented in Sections~\ref{ssec:ubench:ssd} and
\ref{ssec:eval:mysql}, both against the raw device and against F2FS. By using a
separate controller tailored to each application, the dual controller setup
realizes slightly higher performance than the default single LSA controller
setup with 4KiB page size. More importantly, it does so at a significantly
lower overhead, both in terms of DRAM (60\%) and storage capacity (71\%).

Moreover, the dual controller configuration provides segregation of the different 
applications' data, as each controller appends data to separate segments (by
using separate allocators~\Fig{fig:salsa-arch}). This data segregation
allows the dual-controller configuration to perfectly separate the data of the
object store from the data of the MySQL database. The result is a relocation
traffic that is reduced by 28\% compared to the single-controller configuration. 
In this particular case, this reduction does not translate to significant bottom line performance improvement, 
because relocations comprise a small component of the total write workload (7\% for
the single controller setup) which is expected considering that most of the
write workload is sequential (object \pr{PUT}s). Furthermore, the SSD we use
does not offer control over the write streams to the host. Such control, e.g., 
in the form of the Write Streams Directive introduced in the latest version of the NVMe interface~\cite{nvme1.3},
would substantially increase the benefit from stream separation at the host TL level.

\begin{table}
    \footnotesize
\begin{tabular}{|l|c|c|c|c|}\hline
   &  raw    &  F2FS& salsa-single & salsa-dual   \\\hline
\rowcolor[gray]{.9}
sysbench (tps)        & 20.3 & 20.4 & 34.05 & 35.90                 \\
Swift \pr{PUT} (MiB/s)  & 25.5 & 34.28  & 37.29  & 38.19                 \\
\rowcolor[gray]{.9}
DRAM overhead (GiB)   & NA  & 0.85 & 1.66 & 0.68  \\
MD overhead (GiB)     & NA  & 2.11 & 1.82 & 0.53  \\
\rowcolor[gray]{.9}
Relocations (MiB/s)   & NA  & NA & 2.48 & 1.78  \\
\hline
\end{tabular}
\caption{Mixed workload results over raw device (\emph{raw}), over the F2FS filesystem (\emph{F2FS}), \Sys with 1 controller (\emph{salsa-single}) and \Sys with 2 controllers (\emph{salsa-dual}): sysbench throughput in transactions per second (\emph{tps}), Swift object server \pr{PUT} throughput, DRAM overhead, Metadata (\emph{MD}) capacity overhead, and relocation traffic.}
\label{tab:eval:dual-controller}
\end{table}

\subsection{Video server with SMRs and SSDs}
\label{sec:eval:vidserver}

\Sys also supports controllers that combine different storage media.
Here, we evaluate the benefits of running a server for user-generated videos on a \Sys hybrid
controller that combines SSD and SMR drives.
We compare three configurations: using the raw SMR device, using the SMR device
over \Sys, and using a \Sys hybrid controller that employs both a Flash drive
and an SMR drive as described in \Sec{ssec:sys:hybrid}.

We use an XFS filesystem on the device, and
we generate the workload using Filebench \cite{tarasov16:filebench}.
Filebench includes a video-server macro-workload that splits files into two
sets: an active and a passive set. Active files are read from 48
threads, and a file from the passive set is replaced every 10 seconds. 
User-generated videos are typically short, with an average size close to
10MiB\cite{cheng13:youtube}. Furthermore, videos are virtually never deleted,
and most views happen on a relatively small subset of the stored videos.
Subsequently, we modify the workload to use smaller files (10MiB), create new
files instead of replacing files from the passive set every 1 second, and use
direct IO for reads to avoid cache effects.

We run the benchmark for
20 min and show the throughput as reported by Filebench on
\Tab{tab:eval:videoserver}. The write throughput remains at 10~MiB/s for all
cases since we are writing a 10MiB file every second. Using \Sys over the SMR
drive delivers a higher read throughput (6.2 MiB/s versus 4.8 MiB/s) because the periodical writes are less
obstructive to the reads. The hybrid controller
achieves a much higher read throughput of 118.5~MiB/s
by using an SSD to hold the ``hot'' files.

\begin{table}
\centering
\begin{tabular}{|l|r|r|}\hline
throughput   &  R (MiB/s)    &  W (MiB/s)   \\\hline
\rowcolor[gray]{.9}
raw          &        4.8    &                 10.1  \\
salsa        &        6.2    &                 10.1  \\
\rowcolor[gray]{.9}
salsa-hybrid &        118.5  &                 10.0  \\
\hline
\end{tabular}
\caption{Read (\emph{R)} and write (\emph{W}) throughput of video server
macro-benchmark workload.}
\label{tab:eval:videoserver}
\end{table}

\begin{figure}[htb]
    \centering
    \includegraphics[width=\linewidth]{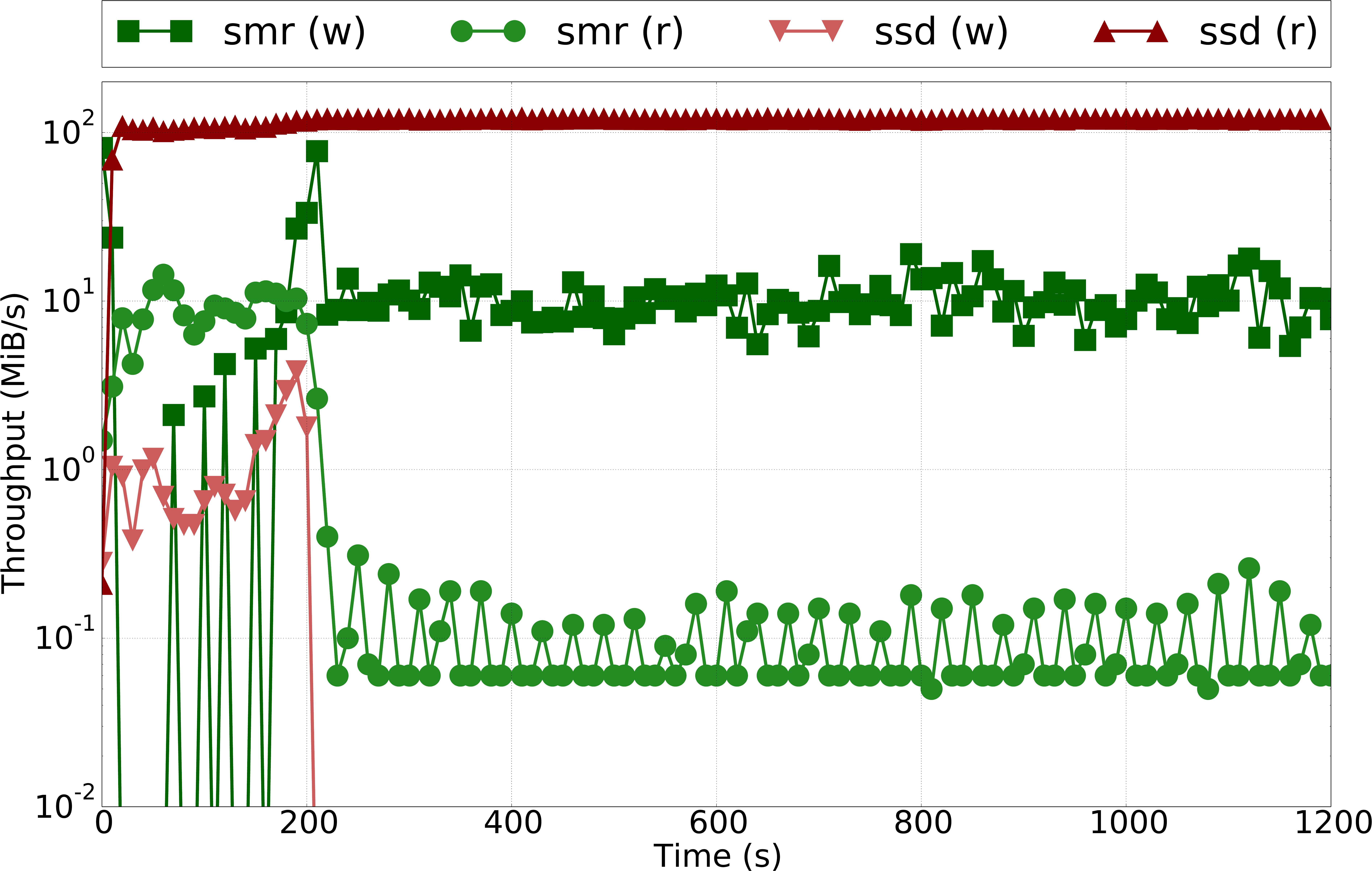}
    \caption{Read and write throughput of the SSD and SMR drive for the video
    server macro-benchmark when using the \Sys hybrid controller.}
    \label{fig:eval:videoserver}
\end{figure}

\Fig{fig:eval:videoserver} gives more insight on the operation of the hybrid
controller by showing the read and write throughput of the SSD and SMR drives as
reported by iostat for the duration of the benchmark (we use a logarithmic scale
on the y axis for clarity). Initially, all files are on the SMR drive. As the
active videos are accessed by the reader threads, data migrates to the SSD and
we observe SSD writes. After about 200 seconds, we reach stable state where all
active videos are in the SSD. At this point, writes are served by the SMR and
reads by the SSD.

%% file: related.tex
\section{Related work}
\label{sec:related}

The log-structured filesystem design was proposed independently of SSDs, as a way
to increase write bandwidth~\cite{rosenblum92:logfs}. Subsequent research work
has proposed Flash-tailored log-structured filesystems to increase performance
either on top of an FTL \cite{f2fs:2015,min:2012:sfs,nilfs06} or by accessing
Flash memory directly~\cite{josephson:2010:dfs,zhang:atc16:parafs}.
Menon introduces log-structured arrays implemented in the storage controller, as
an alternative to RAID\cite{menon95:lsa}. The same approach is followed in Purity
for an enterprise all-Flash array\cite{purity}. All the above systems adopt
append-only writes as a way to minimize random writes on Flash and increase
performance. In our work, we follow a similar approach, but we tailor it to
low-cost commodity devices, while also supporting multiple storage types.

A number of works have identified the limitations of SSD drive TLs, proposing
offloading functionality to the host. 
Jeong et al.\cite{jeong17} propose caching the address mapping table in host
memory, illustrating the problems of limited drive controller resources.
The Virtual Flash Storage Layer
(VFSL)\cite{josephson:2010:dfs,fusionio:ftl:2011,fusionio:apps:2012} is an
attempt to place the TL on the host, exporting a large, virtual block address
space that enables building Flash-friendly applications\cite{marmol14:nvmkv}.
LSDM~\cite{zuck:ngmast14:lsdm} is a host log-structured TL that targets
low-cost SSDs.
Recently, Linux introduced a TL for zoned drives~\cite{dmzoned}
that exclusively targets zoned storage types (e.g., HA or HM SMR).
While our motivation is common with these host TLs, \Sys is fundamentally
different from in two ways.  First, \Sys is designed to support multiple
storage types and devices, using a common codebase to operate on them.
Second, the aforementioned works implement a single TL layer which all
applications use. In \Sys, contrarily, we concede that no single TL
implementation is best for all cases. Instead, \Sys allows for multiple TL
implementation instances (resulting in multiple volumes, for example) on top of
a common \SCM layer.

In a similar spirit, recent attempts expose the internal storage complexities
(e.g., Flash channels\cite{ouyang14:sdf,lightnvme:fast17}, or \GC
controls\cite{ocz2015-hm}) to enable host software to make more intelligent
decisions and reduce controller costs. We view these efforts as orthogonal to
ours: \Sys can operate on and benefit from these drives, but does not depend on
them. Similarly, we view attempts to redefine the interface between
applications and idiosyncratic
storage\cite{ouyang11,zhang:fast12:nameless-writes,amf:fast16,object-ftl:fast13,zhang:atc16:parafs}
also as orthogonal.  Currently, \Sys controllers offer a traditional interface
because we target unmodified applications. Improved interfaces can be
implemented (and co-exist) by individual controllers.

A hybrid system with Flash and disk is presented in\cite{mfdb} for database
storage, where a cost-based model is used to decide which pages to store on
Flash and which pages to store on disk. \Sys is different in that it focuses on
actively transforming the workload to achieve higher performance (and, thus,
lower cost) from the devices using a log-structured approach. A hybrid approach
that we have not investigated is realized by
Griffin~\cite{soundararajan:fast10:griffin} that uses HDDs as a write-cache for SSDs.
Another hybrid approach is taken by Gecko~\cite{shin:fast13:gecko}, where a
log-structured array on top of HDDs in a single TL layer is implemented,
augmented by RAM- and SSD-based caching. \Sys, on the other hand, operates on
SSDs and SMRs, does not rely on data caching, and supports multiple TL
implementation instances.

%% file: conclusion.tex
\section{Conclusion and future work}
\label{sec:conclusions}

In this paper we presented \Sys, a log-structured host TL that
that can be used transparently from applications and offers significant
performance and durability benefits for SSDs and SMR drives.

While we focus on SSDs and SMRs due to their idiosyncrasies in this paper, we
believe that \Sys is also useful for other types of storage. On one hand, a
log-structured TL has significant benefits even in non-idiosyncratic storage
like DRAM\cite{logdram:rumble14} or non-volatile
memory\cite{btt,volos:mnemosyne11}.
On the other hand, coupled with proper policies, a host TL like \Sys can enable
smart data movement between different storage types. We plan to expand on these
ideas in future work. Moreover, in ongoing work we explore building \Sys-native
applications
that execute \Sys as a library in user-space. Among other benefits, this allows
avoiding kernel overheads by utilizing user-space I/O drivers such as the Storage Performance Development Kit (SPDK)~\cite{spdk}.




%% file: ms.bbl
\begin{thebibliography}{10}

\bibitem{skylight:fast15}
{\sc Aghayev, A., and Desnoyers, P.}
\newblock Skylight{\textemdash}a window on shingled disk operation.
\newblock In {\em 13th USENIX Conference on File and Storage Technologies (FAST
  15)\/} (Feb. 2015), pp.~135--149.

\bibitem{aghayev:fast17}
{\sc Aghayev, A., Ts{\textquoteright}o, T., Gibson, G., and Desnoyers, P.}
\newblock Evolving ext4 for shingled disks.
\newblock In {\em 15th {USENIX} Conference on File and Storage Technologies
  ({FAST} 17)\/} (Santa Clara, CA, 2017), {USENIX} Association, pp.~105--120.

\bibitem{andersen10:flash-dc}
{\sc Andersen, D.~G., and Swanson, S.}
\newblock Rethinking flash in the data center.
\newblock {\em IEEE Micro 30}, 4 (July 2010), 52--54.

\bibitem{fusionio:apps:2012}
{\sc Batwara, A.}
\newblock Leveraging flash translation layers for application acceleration.
\newblock Flash Memory Summit, 2012.
\newblock
  \url{http://www.flashmemorysummit.com/English/Collaterals/Proceedings/2012/20120821_TB11_Batwara.pdf}.

\bibitem{lightnvme:fast17}
{\sc Bj{\o}rling, M., Gonzalez, J., and Bonnet, P.}
\newblock Lightnvm: The linux open-channel {SSD} subsystem.
\newblock In {\em 15th {USENIX} Conference on File and Storage Technologies
  ({FAST} 17)\/} (Santa Clara, CA, 2017), {USENIX} Association, pp.~359--374.

\bibitem{google-disks-dc:2016}
{\sc Brewer, E., Ying, L., Greenfield, L., Cypher, R., and T'so, T.}
\newblock Disks for data centers.
\newblock Tech. rep., Google, 2016.

\bibitem{btt}
Btt - block translation table.
\newblock \url{https://www.kernel.org/doc/Documentation/nvdimm/btt.txt}.

\bibitem{andromeda17nvmw}
{\sc Caulfield, L., Xing, M., Tan, Z., and Alexander, R.}
\newblock Andromeda: Building the next-generation high-density storage
  interface for successful adoption.
\newblock \url{http://nvmw.eng.ucsd.edu/2017/assets/slides/51/}, 2017.

\bibitem{Chang2004tecs}
{\sc Chang, L.-P., Kuo, T.-W., and Lo, S.-W.}
\newblock Real-time garbage collection for flash-memory storage systems of
  real-time embedded systems.
\newblock {\em ACM Trans. Embed. Comput. Syst. 3}, 4 (Nov. 2004), 837--863.

\bibitem{cheng13:youtube}
{\sc Cheng, X., Liu, J., and Dale, C.}
\newblock Understanding the characteristics of internet short video sharing: A
  youtube-based measurement study.
\newblock {\em Trans. Multi. 15}, 5 (Aug. 2013), 1184--1194.

\bibitem{chung09:ftl}
{\sc Chung, T.-S., Park, D.-J., Park, S., Lee, D.-H., Lee, S.-W., and Song,
  H.-J.}
\newblock A survey of flash translation layer.
\newblock {\em J. Syst. Archit. 55}, 5-6 (May 2009), 332--343.

\bibitem{purity}
{\sc Colgrove, J., Davis, J.~D., Hayes, J., Miller, E.~L., Sandvig, C., Sears,
  R., Tamches, A., Vachharajani, N., and Wang, F.}
\newblock Purity: Building fast, highly-available enterprise flash storage from
  commodity components.
\newblock In {\em Proceedings of the 2015 ACM SIGMOD International Conference
  on Management of Data\/} (New York, NY, USA, 2015), SIGMOD '15, ACM,
  pp.~1683--1694.

\bibitem{debnath10:flashstore}
{\sc Debnath, B., Sengupta, S., and Li, J.}
\newblock Flashstore: High throughput persistent key-value store.
\newblock {\em Proc. VLDB Endow. 3}, 1-2 (Sept. 2010), 1414--1425.

\bibitem{debnath11:skimpystash}
{\sc Debnath, B., Sengupta, S., and Li, J.}
\newblock Skimpystash: Ram space skimpy key-value store on flash-based storage.
\newblock In {\em Proceedings of the 2011 ACM SIGMOD International Conference
  on Management of Data\/} (2011), SIGMOD '11, pp.~25--36.

\bibitem{dmzoned}
dm-zoned: Zoned block device support.
\newblock
  \url{https://www.kernel.org/doc/Documentation/device-mapper/dm-zoned.txt}.

\bibitem{feldman2013shingled}
{\sc Feldman, T., and Gibson, G.}
\newblock Shingled magnetic recording areal density increase requires new data
  management.
\newblock {\em USENIX; login: Magazine 38}, 3 (2013).

\bibitem{fusionio:ftl:2011}
{\sc {Gary Orenstein}}.
\newblock Optimizing {I/O} operations via the flash translation layer.
\newblock Flash Memory Summit, 2011.
\newblock
  \url{http://www.flashmemorysummit.com/English/Collaterals/Proceedings/2011/20110809_F1B_Orenstein.pdf}.

\bibitem{iostat}
{\sc Godard, S.}
\newblock {\em iostat(1) Linux User's Manual}, July 2013.

\bibitem{gupta09:dftl}
{\sc Gupta, A., Kim, Y., and Urgaonkar, B.}
\newblock {DFTL}: A flash translation layer employing demand-based selective
  caching of page-level address mappings.
\newblock In {\em Proceedings of the 14th International Conference on
  Architectural Support for Programming Languages and Operating Systems\/}
  (2009), ASPLOS XIV, pp.~229--240.

\bibitem{ha10:manual}
{\sc {HGST}}.
\newblock {\em {HGST Ultrastar Archive Ha10}, Hard disk drive specifications},
  June 2015.
\newblock Revision 1.0.

\bibitem{t10-zbc}
{\sc {INCITS} {T10} Technical Committee}.
\newblock {\em Information technology -- {Zoned Block Commands (ZBC)}}, Nov.
  2014.
\newblock Working Draft, Revision 3. Available from
  \url{http://www.t10.org/drafts.htm}.

\bibitem{jeong17}
{\sc Jeong, W., Cho, H., Lee, Y., Lee, J., Yoon, S., Hwang, J., and Lee, D.}
\newblock Improving flash storage performance by caching address mapping table
  in host memory.
\newblock In {\em 9th {USENIX} Workshop on Hot Topics in Storage and File
  Systems (HotStorage 17)\/} (Santa Clara, CA, 2017), {USENIX} Association.

\bibitem{josephson:2010:dfs}
{\sc Josephson, W.~K., Bongo, L.~A., Li, K., and Flynn, D.}
\newblock {DFS}: A file system for virtualized flash storage.
\newblock {\em Trans. Storage 6}, 3 (Sept. 2010), 14:1--14:25.

\bibitem{kim08:bplru}
{\sc Kim, H., and Ahn, S.}
\newblock {BPLRU}: A buffer management scheme for improving random writes in
  flash storage.
\newblock In {\em Proceedings of the 6th USENIX Conference on File and Storage
  Technologies\/} (2008), FAST'08, pp.~16:1--16:14.

\bibitem{knipple:flash17}
{\sc Knipple, S.}
\newblock Leveraging the latest flash in the data center.
\newblock Flash Memory Summit, 2017.
\newblock
  \url{https://www.flashmemorysummit.com/English/Collaterals/Proceedings/2017/20170809_FG21_Knipple.pdf}.

\bibitem{mfdb}
{\sc Koltsidas, I., and Viglas, S.~D.}
\newblock Flashing up the storage layer.
\newblock {\em Proc. VLDB Endow. 1}, 1 (Aug. 2008), 514--525.

\bibitem{nilfs06}
{\sc Konishi, R., Amagai, Y., Sato, K., Hifumi, H., Kihara, S., and Moriai, S.}
\newblock The linux implementation of a log-structured file system.
\newblock {\em SIGOPS Oper. Syst. Rev. 40}, 3 (July 2006), 102--107.

\bibitem{sysbench}
{\sc Kopytov, A.}
\newblock {SysBench: a system performance benchmark 0.5}.
\newblock \url{https://code.launchpad.net/~sysbench-developers/sysbench/0.5}.

\bibitem{f2fs:2015}
{\sc Lee, C., Sim, D., Hwang, J., and Cho, S.}
\newblock {F2FS}: A new file system for flash storage.
\newblock In {\em 13th USENIX Conference on File and Storage Technologies (FAST
  15)\/} (Feb. 2015), pp.~273--286.

\bibitem{amf:fast16}
{\sc Lee, S., Liu, M., Jun, S., Xu, S., Kim, J., and Arvind}.
\newblock Application-managed flash.
\newblock In {\em 14th USENIX Conference on File and Storage Technologies (FAST
  16)\/} (2016), pp.~339--353.

\bibitem{object-ftl:fast13}
{\sc Lu, Y., Shu, J., and Zheng, W.}
\newblock Extending the lifetime of flash-based storage through reducing write
  amplification from file systems.
\newblock In {\em 11th USENIX Conference on File and Storage Technologies (FAST
  13)\/} (2013), pp.~257--270.

\bibitem{ma14:ftl-survey}
{\sc Ma, D., Feng, J., and Li, G.}
\newblock A survey of address translation technologies for flash memories.
\newblock {\em ACM Computing Surveys (CSUR) 46}, 3 (2014), 36.

\bibitem{fbtimeline}
{\sc Mack, R.}
\newblock Building timeline: Scaling up to hold your life story.
\newblock
  \url{https://code.facebook.com/posts/371094539682814/building-timeline-scaling-up-to-hold-your-life-story/}.

\bibitem{marmol14:nvmkv}
{\sc Marmol, L., Sundararaman, S., Talagala, N., Rangaswami, R., Devendrappa,
  S., Ramsundar, B., and Ganesan, S.}
\newblock {NVMKV}: A scalable and lightweight flash aware key-value store.
\newblock In {\em 6th USENIX Workshop on Hot Topics in Storage and File Systems
  (HotStorage 14)\/} (June 2014).

\bibitem{md}
{\em md: Multiple Device driver aka Linux Software {RAID}}.

\bibitem{menon95:lsa}
{\sc Menon, J.}
\newblock A performance comparison of raid-5 and log-structured arrays.
\newblock In {\em High Performance Distributed Computing, 1995., Proceedings of
  the Fourth IEEE International Symposium on\/} (1995), pp.~167--178.

\bibitem{min16atc}
{\sc Min, C., Kashyap, S., Maass, S., and Kim, T.}
\newblock Understanding manycore scalability of file systems.
\newblock In {\em 2016 {USENIX} Annual Technical Conference ({USENIX} {ATC}
  16)\/} (Denver, CO, 2016), {USENIX} Association, pp.~71--85.

\bibitem{min:2012:sfs}
{\sc Min, C., Kim, K., Cho, H., Lee, S.-W., and Eom, Y.~I.}
\newblock Sfs: Random write considered harmful in solid state drives.
\newblock In {\em Proceedings of the 10th USENIX Conference on File and Storage
  Technologies\/} (2012), FAST'12, pp.~12--12.

\bibitem{nanavati15:nvm}
{\sc Nanavati, M., Schwarzkopf, M., Wires, J., and Warfield, A.}
\newblock Non-volatile storage.
\newblock {\em Commun. ACM 59}, 1 (Dec. 2015), 56--63.

\bibitem{nvme1.3}
{N}on-{V}olatile {M}emory {E}xpress ({NVM}e) 1.3.
\newblock \url{http://nvmexpress.org/}.

\bibitem{ocz2015-hm}
{OCZ} announces first {SATA} host managed {SSD}: {Saber} 1000 {HMS}.
\newblock
  \url{http://www.anandtech.com/show/9720/ocz-announces-first-sata-host-managed-ssd-saber-1000-hms}.

\bibitem{ocz-aerospike}
{\sc {OCZ}}.
\newblock Saber 1000 {HMS} series: Performance test report using {Aerospike} db
  and the {YCSB} benchmark tool.

\bibitem{ouyang14:sdf}
{\sc Ouyang, J., Lin, S., Jiang, S., Hou, Z., Wang, Y., and Wang, Y.}
\newblock {SDF}: Software-defined flash for web-scale internet storage systems.
\newblock In {\em Proceedings of the 19th International Conference on
  Architectural Support for Programming Languages and Operating Systems\/}
  (2014), ASPLOS '14, pp.~471--484.

\bibitem{ouyang11}
{\sc Ouyang, X., Nellans, D., Wipfel, R., Flynn, D., and Panda, D.~K.}
\newblock Beyond block {I/O}: Rethinking traditional storage primitives.
\newblock In {\em 2011 IEEE 17th International Symposium on High Performance
  Computer Architecture\/} (Feb 2011), pp.~301--311.

\bibitem{park06:cflru}
{\sc Park, S.-y., Jung, D., Kang, J.-u., Kim, J.-s., and Lee, J.}
\newblock {CFLRU}: A replacement algorithm for flash memory.
\newblock In {\em Proceedings of the 2006 International Conference on
  Compilers, Architecture and Synthesis for Embedded Systems\/} (2006), CASES
  '06, pp.~234--241.

\bibitem{pitchumani15:smrdb}
{\sc Pitchumani, R., Hughes, J., and Miller, E.~L.}
\newblock {SMRDB:} key-value data store for shingled magnetic recording disks.
\newblock In {\em Proceedings of the 8th ACM International Systems and Storage
  Conference\/} (2015), SYSTOR '15, pp.~18:1--18:11.

\bibitem{rosenblum92:logfs}
{\sc Rosenblum, M., and Ousterhout, J.~K.}
\newblock The design and implementation of a log-structured file system.
\newblock {\em ACM Transactions on Computer Systems (TOCS) 10}, 1 (1992),
  26--52.

\bibitem{logdram:rumble14}
{\sc Rumble, S.~M., Kejriwal, A., and Ousterhout, J.}
\newblock Log-structured memory for {DRAM}-based storage.
\newblock In {\em Proceedings of the 12th USENIX Conference on File and Storage
  Technologies (FAST 14)\/} (Santa Clara, CA, 2014), USENIX, pp.~1--16.

\bibitem{saxena10:fvm}
{\sc Saxena, M., and Swift, M.~M.}
\newblock {FlashVM}: Virtual memory management on flash.
\newblock In {\em Proceedings of the 2010 USENIX Conference on USENIX Annual
  Technical Conference\/} (2010), USENIXATC'10, pp.~14--14.

\bibitem{saxena12:flashtier}
{\sc Saxena, M., Swift, M.~M., and Zhang, Y.}
\newblock {FlashTier:} a lightweight, consistent and durable storage cache.
\newblock In {\em Proceedings of the 7th ACM European Conference on Computer
  Systems\/} (2012), EuroSys '12, pp.~267--280.

\bibitem{google:flash:fast16}
{\sc Schroeder, B., Lagisetty, R., and Merchant, A.}
\newblock Flash reliability in production: The expected and the unexpected.
\newblock In {\em 14th {USENIX} Conference on File and Storage Technologies
  ({FAST} 16)\/} (Santa Clara, CA, 2016), {USENIX} Association, pp.~67--80.

\bibitem{st8:manual}
{\sc Seagate}.
\newblock {\em Archive HDD: v2 {SATA} Product Manual: ST8000AS0022,
  ST6000AS0022}, Nov. 2015.
\newblock Revision F.

\bibitem{onestor}
{\sc {Seagate Corporation}}.
\newblock {\em {OneStor} {AP-2584} Datasheet}.

\bibitem{seagete:sshd}
Seagate: Solid state hybrid technology.
\newblock \url{http://www.seagate.com/solutions/solid-state-hybrid/products/}.

\bibitem{shin:fast13:gecko}
{\sc Shin, J.~Y., Balakrishnan, M., Marian, T., and Weatherspoon, H.}
\newblock Gecko: Contention-oblivious disk arrays for cloud storage.
\newblock In {\em Presented as part of the 11th {USENIX} Conference on File and
  Storage Technologies ({FAST} 13)\/} (San Jose, CA, 2013), {USENIX},
  pp.~285--297.

\bibitem{soundararajan:fast10:griffin}
{\sc Soundararajan, G., Prabhakaran, V., Balakrishnan, M., and Wobber, T.}
\newblock Extending ssd lifetimes with disk-based write caches.
\newblock In {\em 8th USENIX Conference on File and Storage Technologies\/}
  (2010), FAST'10, pp.~8--8.

\bibitem{spdk}
Storage performance development kit.
\newblock \url{http://www.spdk.io/}.

\bibitem{radu-freq}
{\sc Stoica, R., and Ailamaki, A.}
\newblock Improving flash write performance by using update frequency.
\newblock {\em Proc. VLDB Endow. 6}, 9 (July 2013), 733--744.

\bibitem{openstack-swift}
{OpenStack} {Swift}.
\newblock \url{http://swift.openstack.org/}.

\bibitem{openstack-swift-arch}
{Swift} architectural overview.
\newblock
  \url{http://docs.openstack.org/developer/swift/overview_architecture.html}.

\bibitem{openstack-sw-conf}
{Swift} software configuration procedures.
\newblock
  \url{http://docs.openstack.org/developer/swift/ops_runbook/procedures.html}.

\bibitem{ripq:fast15}
{\sc Tang, L., Huang, Q., Lloyd, W., Kumar, S., and Li, K.}
\newblock {RIPQ}: Advanced photo caching on flash for facebook.
\newblock In {\em 13th USENIX Conference on File and Storage Technologies (FAST
  15)\/} (Feb. 2015), pp.~373--386.

\bibitem{tarasov16:filebench}
{\sc Tarasov, V., Zadok, E., and Shepler, S.}
\newblock Filebench: A flexible framework for file system benchmarking.
\newblock {\em ;login 41}, 1 (2016).

\bibitem{Trivedi17flashnet}
{\sc Trivedi, A., Ioannou, N., Metzler, B., Stuedi, P., Pfefferle, J.,
  Koltsidas, I., Kourtis, K., and Gross, T.~R.}
\newblock Flashnet: Flash/network stack co-design.
\newblock In {\em Proceedings of the 10th ACM International Systems and Storage
  Conference\/} (New York, NY, USA, 2017), SYSTOR '17, ACM, pp.~15:1--15:14.

\bibitem{idc-flash}
{\sc Villars, R.~L., and Burgener, E.}
\newblock {IDC}: Building data centers for today’s data driven economy: The
  role of flash.
\newblock
  \url{https://www.sandisk.com/business/datacenter/resources/white-papers/flash-in-the-data-center-idc},
  July 2014.

\bibitem{volos:mnemosyne11}
{\sc Volos, H., Tack, A.~J., and Swift, M.~M.}
\newblock Mnemosyne: Lightweight persistent memory.
\newblock In {\em Proceedings of the Sixteenth International Conference on
  Architectural Support for Programming Languages and Operating Systems\/}
  (2011), ASPLOS XVI, pp.~91--104.

\bibitem{wang14:lsm-ssd}
{\sc Wang, P., Sun, G., Jiang, S., Ouyang, J., Lin, S., Zhang, C., and Cong,
  J.}
\newblock An efficient design and implementation of lsm-tree based key-value
  store on open-channel ssd.
\newblock In {\em Proceedings of the Ninth European Conference on Computer
  Systems\/} (2014), EuroSys '14, pp.~16:1--16:14.

\bibitem{wdblack2}
{\sc {{Western Digital Technologies}}}.
\newblock {\em {WD} Black2 Dual Drive User Manual}, Nov 2013.

\bibitem{wood:tm09}
{\sc Wood, R., Williams, M., Kavcic, A., and Miles, J.}
\newblock The feasibility of magnetic recording at 10 terabits per square inch
  on conventional media.
\newblock {\em Magnetics, IEEE Transactions on 45}, 2 (Feb 2009), 917--923.

\bibitem{zhang:atc16:parafs}
{\sc Zhang, J., Shu, J., and Lu, Y.}
\newblock Parafs: A log-structured file system to exploit the internal
  parallelism of flash devices.
\newblock In {\em Proceedings of the 2016 USENIX Conference on Usenix Annual
  Technical Conference\/} (2016), USENIX ATC '16, pp.~87--100.

\bibitem{zhang:fast12:nameless-writes}
{\sc Zhang, Y., Arulraj, L.~P., Arpaci-Dusseau, A.~C., and Arpaci-Dusseau,
  R.~H.}
\newblock De-indirection for flash-based ssds with nameless writes.
\newblock In {\em Proceedings of the 10th USENIX Conference on File and Storage
  Technologies\/} (2012), FAST'12, pp.~1--1.

\bibitem{zheng13:cosbench}
{\sc Zheng, Q., Chen, H., Wang, Y., Zhang, J., and Duan, J.}
\newblock Cosbench: Cloud object storage benchmark.
\newblock In {\em Proceedings of the 4th ACM/SPEC International Conference on
  Performance Engineering\/} (2013), ICPE '13, pp.~199--210.

\bibitem{zuck:ngmast14:lsdm}
{\sc Zuck, A., Kishon, O., and Toledo, S.}
\newblock {LSDM:} improving the performance of mobile storage with a
  log-structured address remapping device driver.
\newblock In {\em Proceedings of the 2014 Eighth International Conference on
  Next Generation Mobile Apps, Services and Technologies\/} (Washington, DC,
  USA, 2014), NGMAST '14, IEEE Computer Society, pp.~221--228.

\end{thebibliography}
